\documentclass[prd,showpacs,epsfig,onecolumn]{revtex4}
\usepackage{psfrag}
\usepackage{graphicx}
\usepackage{dcolumn}
\usepackage{bm}

\begin{document}

\title{Analyzing weak lensing of the cosmic microwave background using
the likelihood function}

\author{Christopher M. Hirata}
\email{chirata@princeton.edu}

\author{Uro\v s Seljak}
\email{useljak@princeton.edu}
\affiliation{Physics Department, Princeton University, Princeton, NJ 08544}

\date{September 24, 2002}

\begin{abstract}
Future experiments will produce high-resolution temperature maps of the
cosmic microwave background (CMB) and are expected to reveal the signature
of gravitational lensing by intervening large-scale structures.  We
construct all-sky maximum-likelihood estimators that use the lensing
effect to estimate the projected density (convergence) of these
structures, its power spectrum, and cross-correlation with other
observables.  This contrasts with earlier quadratic-estimator approaches
that Taylor-expanded the observed CMB temperature to linear order in the
lensing deflection angle; these approaches gave estimators for the
temperature-convergence correlation in terms of the CMB three-point
correlation function and for the convergence power spectrum in terms of
the CMB four-point correlation
function, which can be biased and non-optimal due to terms beyond the
linear order.  
We show that for sufficiently weak lensing, the maximum-likelihood
estimator reduces to the computationally less
demanding quadratic estimator.  The maximum likelihood and quadratic
approaches are compared by evaluating the
root-mean-square (RMS) error and bias in the reconstructed convergence
map in a numerical simulation; 
it is found that both the RMS errors and bias 
are of order 1 percent for the case of Planck and of order 10--20 percent
for a 1 arcminute beam 
experiment.  We
conclude that for recovering lensing information from temperature data
acquired by these experiments, the quadratic
estimator is close to optimal, but
further work will be required to determine whether this is also the case
for lensing of the CMB polarization field.
\end{abstract}


\maketitle

\section{\label{sec:s1}Introduction}

Gravitational weak lensing of the cosmic microwave background (CMB) has
been recognized as a potential indicator of
large-scale structure in the universe.  Compared to galaxy surveys, weak
lensing has the advantage of directly tracing
the matter density, thus avoiding the uncertainties associated with the
relationship between the distributions of
galaxies and of mass \cite{1999PhRvL..82.2636S}.  Because the CMB is the
most distant background object that can be used
for weak lensing studies, it probes the matter distribution at higher
redshifts than can be reached by galaxy weak
lensing and is sensitive to the largest observable scales in the universe
 \cite{1999PhRvL..82.2636S,1999PhRvD..59l3507Z,2001PhRvD..63d3501B,
2001ApJ...557L..79H,2000PhRvD..62d3007H}.

In addition to providing data on the power spectrum of density
fluctuations on these large scales, CMB weak lensing may yield
constraints on the expansion history of the
universe by making possible a measurement of the integrated Sachs-Wolfe
(ISW) effect.  The ISW effect (the change in temperature of the CMB
radiation as it passed through
a changing gravitational potential) is smaller than the primary CMB
fluctuations produced in the early universe and consequently can be
detected only through the
cross-correlation of CMB observations with some tracer of the
gravitational potential.  Because it is sensitive directly to the
potential, weak lensing is an ideal
candidate for this cross-correlation
\cite{1999PhRvD..60d3504S,1999PhRvD..59j3002G}.

Because detection of CMB weak lensing may be possible with near-future
satellite experiments, such as Planck and
possibly even MAP, several algorithms have been proposed for estimating
matter distributions, power spectra, and ISW
cross-correlations from CMB temperature maps.  Some of these methods are
based on local statistics, such as the products
of gradients of the temperature field \cite{1999PhRvL..82.2636S}.  Recently Hu
\cite{2000PhRvD..62d3007H,2001ApJ...557L..79H}, working to linear order
in the deflection angle, determined the optimal
quadratic estimator (i.e. quadratic in the CMB temperature map) for the
deflection field.  Within this linear
approximation, the corresponding power spectrum estimator makes full use
of the information in the CMB four-point
correlation function
\cite{2000PhRvD..62f3510Z,2001PhRvD..64h3005H}.  However, the limits to
the validity of the linear order approximation
have not been well-determined, and the possibility of obtaining more
information on lensing from higher-order
correlation functions has not been studied in detail. Neglect of
nonlinear terms may also create a bias in the
quadratic estimators of the power spectrum. The nonlinear terms 
may be important whenever the deflection angle is comparable to the scale 
of CMB fluctuation used in the reconstruction of lensing potential. 
The deflection angle is of the order of several 
arcminutes and for high resolution experiments significant amount of 
lensing information comes from CMB modes on the same scale, indicating that the 
nonlinear terms may be important.  
In order to address these issues, we use the likelihood function to
construct estimators rather than assuming an estimator with a particular
form (local, quadratic,
etc.) and avoid linearizing in the deflection field except to compare our
results to previous work and where necessary for computational tractability.

We work principally in position space rather than harmonic space.  This
is done partly because real data are obtained in
position space, and partly to show how the harmonic-space estimators
\cite{2001ApJ...557L..79H} can be derived from
position-space arguments; also, the generalization of the position-space
analysis to anisotropic instrument noise
is more transparent.  We also do not consider the reconstruction of
matter
distributions from CMB
polarization; although polarization can theoretically yield much better
information about lensing than CMB temperature
fluctuations \cite{2002ApJ...574..566H}, it is also computationally more
demanding, so we defer a more
careful analysis to a future work. 

We will proceed as follows: Section \ref{sec:s2} introduces our formalism
and notation, and defines the basic
mathematical operations that will be used in the rest of the
paper.  Section \ref{sec:s3} considers the likelihood
function for the CMB and its dependence on the lensing potential (the
potential that generates the deflection
field).  In Section \ref{sec:s4} we consider the maximum likelihood
estimators for the power spectrum of the lensing
potential and its cross-correlation with the CMB.  In Section
\ref{sec:s5}, we describe our numerical implementation of
the estimators from Sections \ref{sec:s3} and \ref{sec:s4}; the
performance of the estimators, as determined
numerically, is described in Section \ref{sec:s6}.  We conclude in
Section \ref{sec:s7}.

\section{\label{sec:s2}Formalism}

\subsection{\label{sec:s2a}CMB}

The cosmic microwave background temperature fluctuation $\tilde\Theta$ in
a particular direction ${\bf n}$ on the unit sphere is defined by
$\tilde\Theta({\bf n}) = T({\bf n})/T_0-1$ where $T({\bf n})$ is the CMB
temperature in direction ${\bf n}$ and $T_0$=2.72 K is the mean
temperature of the CMB.  This temperature fluctuation can be expressed in
harmonic space as

\begin{equation}
\tilde\Theta({\bf n}) = \sum_{l=0}^\infty \sum_{m=-l}^l \tilde\Theta_{lm}
Y_{lm}({\bf n}),
\label{eq:sphexp}
\end{equation}
where the $Y_{lm}$ are spherical harmonics and $\tilde\Theta_{lm}$ are
the corresponding coefficients.  The spherical harmonics are
orthogonal and are normalized so that their squared amplitude integrates
to one over the sphere: $\int_{S^2} |Y_{lm}^2| d\Omega=1$,
and the
transformation of equation (\ref{eq:sphexp}) can thus be inverted as:

\begin{equation}
\tilde\Theta_{lm} = \int_{S^2} d^2{\bf n} Y_{lm}^\ast({\bf
n}) \tilde\Theta({\bf n}).
\label{eq:sphexp2}
\end{equation}
Because the
statistical average $\langle\tilde\Theta_{lm}\rangle =0$, we extensively
use the power spectrum.  The power spectrum is defined for
a statistically isotropic temperature fluctuation as the variance

\begin{equation}
\langle \tilde\Theta^\ast_{l'm'}\tilde\Theta_{lm}\rangle =
C^{\tilde\Theta\tilde\Theta}_l\delta_{ll'}\delta_{mm'}.
\label{eq:cldef}
\end{equation}
For gravitational lensing work, we distinguish three temperature
fluctuations: the unlensed temperature fluctutation
$\Theta$; the lensed temperature fluctuation $\tilde\Theta$; and the
measured temperature fluctuation $\hat\Theta$.  Throughout this paper, we
will take the primary (unlensed) anisotropy $\Theta$ to be a Gaussian
random field.  The measurement is related to the actual temperature
fluctuation by the instrument noise, $\epsilon$:

\begin{equation}
\hat\Theta({\bf n}) = \tilde\Theta({\bf n}) + \epsilon({\bf n}).
\label{eq:noisedef}
\end{equation}
We assume that the instrument noise $\epsilon$ is independent of
$\tilde\Theta$.

Occasionally we will use the flat-sky approximation, in which a map
$\Theta$ can be expanded in Fourier modes, $\tilde\Theta({\bf n}) =
{1\over\sqrt{4\pi}}
\sum_{\bf l} \tilde\Theta_{\bf l} e^{-i{\bf l}\cdot{\bf n}}$.  The
Fourier modes are normalized over an area of $4\pi$, and populate the
${\bf l}$-plane with a
two-dimensional
density of $1/\pi$; this ensures that the flat-sky and all-sky
normalizations are consistent on small scales.

\subsection{\label{sec:s2b}Lensing}

Gravitational lensing of the CMB by scalar perturbations can be expressed
in terms of the lensing potential $\Phi$,
defined by

\begin{equation}
\tilde\Theta({\bf n}) = \Theta({\bf n} + \nabla\Phi({\bf n})),
\label{eq:potdef}
\end{equation}
where $\nabla$ is the 2-dimensional gradient operator on the unit
sphere.  The lensing potential $\Phi$ is the
projected gravitational potential along the line of sight (see the
Appendix \ref{sec:aphi} for details),

\begin{equation}
\Phi({\bf n}) = -2 \int_0^{r_{ls}} dr \Psi(r{\bf n},-r) \left( {1\over
T(r)} - {1\over T(r_{ls})} \right),
\label{eq:potform}
\end{equation}
where $r_{ls}$ is the comoving distance to the last-scatter surface,
$\Psi({\bf x},\tau)$ is the gravitational scalar
potential at comoving position ${\bf x}$ and conformal time $\tau$, and
$T(r)$ is the tangentlike function ($\tan r$,
$r$, or $\tanh r$ depending on whether the universe is closed, spatially
flat, or open).  The convergence
$\kappa=-{1\over 2}\nabla^2\Phi$ is positive when structures along the
line of sight act as a converging lens (i.e. when they magnify the CMB)
and is negative for a diverging lens.  Conceptually, we would thus expect
$\kappa$ to be
a measure of the projected density perturbation; as shown in the
Appendix, this is indeed the case.  We define the
power spectra $C^{\Phi\Phi}_l$ and
$C^{\kappa\kappa}_l=l^2(l+1)^2C^{\Phi\Phi}_l/4$, and the cross-correlation
$C^{\Theta\Phi}_l$, in analogy to equation (\ref{eq:cldef}).

We will in several instances require use of the lensing operator
$\Lambda$ that performs the operation in equation (\ref{eq:potdef}):

\begin{equation}
\Lambda[\Phi]\Theta({\bf n}) = \Theta({\bf n} + \nabla\Phi({\bf n})).
\label{eq:lambda}
\end{equation}
On
occasion, we shall refer to the linear approximation to the lensing operator:

\begin{equation}
\tilde\Theta = \Lambda\Theta \approx \Theta + \nabla\Theta\cdot\nabla\Phi.
\label{eq:linear}
\end{equation}

Note that we have used $\tilde\Theta$ to represent the lensed CMB
temperature and $\Theta$ to represent the unlensed temperature; some
authors have used this convention
\cite{2000PhRvD..62d3007H}, while others
\cite{1999PhRvL..82.2636S,2002ApJ...574..566H,2001ApJ...557L..79H,
2001PhRvD..64h3005H} have used $\Theta$ for the lensed and
$\tilde\Theta$ for the unlensed temperature.

\subsection{\label{sec:s2c}Convolutions and Integrals}

A convolution of a function $\Theta$ on the unit sphere with kernel $C$
is written as

\begin{equation}
C\Theta({\bf n}) = \int_\Omega d^2{\bf n}' C({\bf n},{\bf
n}') \Theta({\bf n}'),
\label{eq:convdef}
\end{equation}
where $\Omega$ is the region in which we have data, and the kernel $C$
can be decomposed in multipoles using the Legendre polynomials:

\begin{equation}
C({\bf n},{\bf n}') = \sum_{l=0}^\infty {2l+1\over 4\pi} C_l P_l({\bf
n}\cdot{\bf n}').
\label{eq:pl}
\end{equation}
We will also need to take the inverse operation $C^{-1}$ such that
$CC^{-1}\Theta = \Theta$.  In the case of a true
full-sky experiment (i.e. one that acquires usable data over the full
$4\pi$ steradians), the $C^{-1}$ operation is
trivial: we apply a convolution with a $C^{-1}$ kernel with multipoles
$(C^{-1})_l=(C_l)^{-1}$.  The inversion is more
difficult on a portion of the sphere, as discussed in Section
\ref{sec:s5}.

Finally, we make use of the notation derived from linear algebra: our
``column vectors'' are functions on $\Omega$, and our ``matrices'' are
linear operators on this set of functions: $Av({\bf x}) = \int_\Omega
A({\bf x},{\bf y})v({\bf y}) d^2{\bf y}$.  Example uses of this notation
are $u^Tv = \int_\Omega uv d^2{\bf n}$ and $ A^T({\bf x},{\bf y})= A({\bf
y},{\bf x})$.

\section{\label{sec:s3}Likelihood Analysis}

We analyze the likelihood function for gravitational lensing because this
function retains all of the information
provided by the observations.  In particular, we can compare the
``optimal'' maximum likelihood estimators
(MLEs) to previous results.  We examine the relationship between the
quadratic estimators and the likelihood-based
estimators and the criteria for their equivalence, i.e. for optimality of
the quadratic estimator.

We will see that the lensing potential $\Phi$ is detectable because its
presence breaks spherical symmetry and thus
causes correlations between the different spherical harmonic modes of the
temperature field $\tilde\Theta$, i.e. it
creates off-diagonal elements in the
covariance $C^{\tilde\Theta\tilde\Theta}=\langle
\tilde\Theta\tilde\Theta^T\rangle$ when expressed in the spherical
harmonic basis; this is manifested in real space by an anisotropic
correlation function
$C^{\tilde\Theta\tilde\Theta}({\bf x},{\bf y})$. 
Since these off-diagonal elements are, in the linear approximation,
proportional to the lensing
potential $\Phi$, we could take $\tilde\Theta\tilde\Theta^T$ as a crude
estimate of the covariance
$C^{\tilde\Theta\tilde\Theta}$ and form linear combinations of the
off-diagonal elements to construct an estimator for
$\Phi$; this
is the essence of the quadratic estimator methods
\cite{2001ApJ...557L..79H}.  (In the presence of
instrument noise we measure $\hat\Theta$ and not
$\tilde\Theta$ but the idea is the same.)
Note that while some quadratic estimators
(e.g. \cite{1999PhRvL..82.2636S}) have been derived from considering the
magnification and shear of small-scale CMB features by
larger-scale lensing modes, in analogy to the weak lensing of galaxies,
such a picture is not essential to the quadratic estimation framework --
quadratic estimation is
possible whenever the linear approximation to
$C^{\tilde\Theta\tilde\Theta}$ is valid.
 The likelihood method, while somewhat more involved, is useful to
investigate for two reasons: first, unlike quadratic estimators, MLEs are
guaranteed to be asymptotically efficient
(i.e. it is impossible to
achieve lower error than the MLE in the limit of an infinite amount of
data); and second, the likelihood approach
retains its validity even when higher-order [e.g. $O(\Phi^2)$] terms in
the covariance are important.

This section will be organized as follows.  In Section \ref{sec:s3a}, we
introduce the likelihood function and its basic properties and give a
formal expression for
it.  We maximize the likelihood function using the calculus of variations
(Section \ref{sec:s3b}) and proceed to show that within the linear
approximation [equation
(\ref{eq:linear})] the maximum likelihood estimator reduces to the
optimally weighted quadratic estimator (Section \ref{sec:s3c}).  We
examine our ability to reconstruct
the primary CMB
anisotropy $\Theta$ in Section \ref{sec:primaryrec}.  We conclude in
Sections \ref{sec:s3d} and \ref{sec:s3f} by examining the limits of
validity of the linear
approximation.

\subsection{\label{sec:s3a}Likelihood Function}

Likelihood maximization is a generally applicable method to statistical
estimation problems.  A statistical estimation problem involves a data
set, in this case the
measured CMB temperature fluctuation $\hat\Theta({\bf n}_i)$ at N points
$\{ {\bf n}_1,..,{\bf n}_N\}$, which has a probability distribution
determined by a set of
parameters, in this case the values of the lensing potential $\Phi$.  The
problem is to estimate the unknown parameters $\Phi$ from the
observations $\hat\Theta$.  We
represent the probability distribution by a density function $P$, which
is related to the
differential probability $d\Pi$ for obtaining temperature measurements
between $\hat\Theta({\bf n}_i)$ and $\hat\Theta({\bf
n}_i)+d\hat\Theta({\bf n}_i)$:

\begin{equation}
d\Pi = P(\hat\Theta |\Phi) d\hat\Theta({\bf n}_1) ...  d\hat\Theta({\bf n}_N).
\label{eq:pr1}
\end{equation}
The maximum likelihood estimation method simply selects the value of
$\Phi$ that yields the largest value of $P$, i.e. the value of $\Phi$
that would have been most likely
to generate the observed $\hat\Theta$.  While this method is very general
and can be applied to a wide range of problems, maximum likelihood
estimators (MLEs) are
frequently very difficult to compute, as is the case here.

For convenience, we will work not with the likelihood function but with
its negative logarithm ${\cal L}$, which is defined by the relation

\begin{equation}
{\cal L}[\Phi] = -\ln P(\hat\Theta |\Phi).
\label{eq:likelihood}
\end{equation}
If we assume Gaussian instrument noise of covariance
$C^{\epsilon\epsilon}$, we find that for fixed $\Phi$, $\hat\Theta$ is a
Gaussian random field with covariance

\begin{equation}
C^{\hat\Theta \hat\Theta}[\Phi] = \Lambda[\Phi] C^{\Theta\Theta}
\Lambda[\Phi]^T + C^{\epsilon\epsilon},
\label{eq:chat}
\end{equation}
where the transpose $\Lambda^T$ of the linear operator $\Lambda$ is
defined by $\Lambda^T({\bf x},{\bf y})=\Lambda({\bf
y},{\bf x})$.  The probability density of $\hat\Theta$ is then related to
its covariance via the usual relation for a Gaussian:

\begin{equation}
P(\hat\Theta|\Phi) = {1\over (2\pi)^{N/2}\sqrt{\det
C^{\hat\Theta\hat\Theta}}} \exp\left( -{1\over 2}\hat\Theta^T
C^{\hat\Theta\hat\Theta-1}\hat\Theta \right).
\label{eq:pr2}
\end{equation}
Combining this with the definition (\ref{eq:likelihood}) and the standard
Gaussian probability density
formula, we find that

\begin{equation}
{\cal L}[\Phi] = {1\over 2} \hat\Theta^T (C^{\hat\Theta
\hat\Theta}[\Phi])^{-1} \hat\Theta + {1\over 2}\ln\det
C^{\hat\Theta \hat\Theta}[\Phi].
\label{eq:lform}
\end{equation}
In some cases, we will use a Gaussian random prior for $\Phi$, in which
case we will use the negative log posterior
probability ${\cal P}$ in place of the negative log likelihood ${\cal
L}$.  The Gaussian prior for $\Phi$ is

\begin{equation}
P(\Phi|C^{\Phi\Phi}) = {1\over (2\pi)^{N/2} \sqrt{\det C^{\Phi\Phi}}}
\exp \left( -{1\over 2}
\Phi^T C^{\Phi\Phi-1}\Phi\right).
\label{eq:gauss1}
\end{equation}
where $N$ is the number of pixels in the map.
From this the negative log posterior probability can be determined (up to
an irrelevant constant) to be

\begin{equation}
{\cal P}[\Phi; C^{\Phi\Phi}] = {\cal L}[\Phi] - \ln P(\Phi|C^{\Phi\Phi})
= {1\over 2} \hat\Theta^T (C^{\hat\Theta \hat\Theta}[\Phi])^{-1}
\hat\Theta + {1\over 2}\ln\det C^{\hat\Theta
\hat\Theta}[\Phi]
+ {1\over 2} \Phi^T ( C^{\Phi\Phi})^{-1} \Phi + {1\over 2}\ln\det C^{\Phi\Phi},
\label{eq:posterior}
\end{equation}
where $C^{\Phi\Phi}$ is the covariance of the prior for $\Phi$.

\subsection{\label{sec:s3b}Likelihood-Based Estimators}

We construct estimators using ${\cal L}$ and ${\cal P}$ by setting their
functional derivatives with respect to
$\Phi({\bf n})$ equal to zero.  Differentiating equation
(\ref{eq:lform}) gives

\begin{equation}
{\delta{\cal L} [\Phi]\over\delta\Phi} = -{1\over 2} \hat\Theta^T
(C^{\hat\Theta \hat\Theta}[\Phi])^{-1} {\delta C^{\hat\Theta
\hat\Theta}[\Phi] \over\delta\Phi} (C^{\hat\Theta \hat\Theta}[\Phi])^{-1}
\hat\Theta
+ {1\over 2} {\rm Tr} \left[ (C^{\hat\Theta \hat\Theta}[\Phi])^{-1}
{\delta C^{\hat\Theta \hat\Theta}[\Phi]
\over\delta\Phi} \right].
\end{equation}
Using equation ({\ref{eq:chat}), we calculate the functional derivative
of $C^{\hat\Theta \hat\Theta}[\Phi]$:

\begin{equation}
{\delta C^{\hat\Theta \hat\Theta}[\Phi]({\bf y},{\bf
z}) \over\delta\Phi({\bf x})}
=\int_\Omega (\Lambda[\Phi] C^{\Theta\Theta})({\bf y},{\bf
y}') {\delta\Lambda[\Phi]({\bf z},{\bf y}')\over\delta\Phi({\bf x})}
d^2{\bf y}'
+ {\rm transpose}.
\label{eq:lgrad}
\end{equation}
We differentiate $\Lambda$ using equation (\ref{eq:lambda}):

\begin{equation}
{\delta\over\delta\Phi({\bf x})}(\Lambda[\Phi] v)({\bf w})
= \int_\Omega d^2{\bf x}' {\delta\nabla\Phi({\bf x}')\over\delta\Phi({\bf
x})}\cdot
{\delta\over\delta\nabla\Phi({\bf x}')}(\Lambda[\Phi] v)({\bf w})
= (\nabla_{\bf w} \delta^{(2)}({\bf w}-{\bf x})) \cdot
(\Lambda[\Phi]\nabla v)({\bf w}).
\label{eq:c7}
\end{equation}
Using this relation and integration by parts, we convert equation
(\ref{eq:lgrad}) into

\begin{eqnarray}
{\delta C^{\hat\Theta \hat\Theta}[\Phi]({\bf y},{\bf
z}) \over\delta\Phi({\bf x})}
&& = \int_\Omega d^2{\bf y}' (\Lambda[\Phi] C^{\Theta\Theta})({\bf y},{\bf y}')
(\nabla_{\bf z} \delta^{(2)}({\bf z}-{\bf x})) \cdot
(\Lambda[\Phi]\nabla)({\bf z},{\bf y}')
+ {\rm transpose}\nonumber\\*
&& =(\nabla_{\bf z} \delta^{(2)}({\bf z}-{\bf x})) \cdot
(\Lambda[\Phi]\nabla C^{\Theta\Theta}\Lambda[\Phi]^T)({\bf z},{\bf y})
+{\rm transpose}.
\label{eq:c8}
\end{eqnarray}
We also express the trace as an expectation value using the identity
${\rm Tr~}(X) = \langle uXC^{-1}u\rangle$ with $u$
drawn from a Gaussian distribution of covariance $C$, and integrate by
parts again to yield

\begin{eqnarray}
{\delta{\cal L}[\Phi]\over\delta\Phi} = && \nabla\cdot \left[ \hat\Theta
(C^{\hat\Theta \hat\Theta}[\Phi])^{-1} \Lambda[\Phi] \nabla
C^{\Theta\Theta} \Lambda[\Phi]^{-1}
(C^{\hat\Theta \hat\Theta}[\Phi])^{-1} \hat\Theta \right]
\nonumber\\*
&& - \left< \nabla\cdot \left[ \hat\Theta (C^{\hat\Theta
\hat\Theta}[\Phi])^{-1} \Lambda[\Phi] \nabla C^{\Theta\Theta}
\Lambda[\Phi]^{-1} (C^{\hat\Theta 
\hat\Theta}[\Phi])^{-1} \hat\Theta \right] \right>.
\label{eq:c9}
\end{eqnarray}
The functional derivative of ${\cal P}$ differs by the addition of a
$C^{\Phi\Phi-1} \Phi$ term.  The maximum-likelihood estimator
$\hat\Phi$ for $\Phi$ is then the solution to $\delta{\cal
L}/\delta\Phi=0$.  If we then define the likelihood gradient $G[\Phi]$
by:

\begin{eqnarray}
G[\Phi] \equiv {\delta{\cal L}\over\delta\Phi} = && \nabla\cdot \biggl[
(C^{\hat\Theta \hat\Theta}[\hat \Phi]^{-1}
\hat\Theta) \Lambda[\hat\Phi] \nabla C^{\Theta\Theta}
\Lambda[\hat \Phi]^{-1} (C^{\hat\Theta \hat\Theta}[\hat \Phi])^{-1}
\hat\Theta \nonumber\\* &&
- \left< (C^{\hat\Theta \hat\Theta}[\hat
\Phi])^{-1}\hat\Theta) \Lambda[\hat \Phi] \nabla C^{\Theta\Theta}
\Lambda[\hat \Phi]^{-1} (C^{\hat\Theta \hat\Theta}[\hat \Phi])^{-1}
\hat\Theta \right> \biggr],
\label{eq:c10}
\end{eqnarray}
then the maximum likelihood estimator becomes:

\begin{equation}
G[\hat\Phi] =0,
\label{eq:c11}
\end{equation}
whereas the mode of the posterior probability distribution (i.e. maximum
of $e^{-\cal P}$) is the solution $\hat\Phi$ to:

\begin{equation}
\hat \Phi = -C^{\Phi\Phi} G[\hat\Phi].
\label{eq:c12}
\end{equation}
In deriving equation (\ref{eq:c10}), we have dropped boundary terms.  In
our implementation (Section \ref{sec:s5}) we
simply do not work near the survey boundaries, however, the formalism can
be generalized to include these by setting
$C^{\epsilon\epsilon}=\infty$ in the unscanned regions.  This does not
cause numerical difficulties because the inifinte eigenvalues of
$C^{\epsilon\epsilon}$ (and hence $C^{\hat\Theta\hat\Theta}$) become null
eigenvalues of $C^{\hat\Theta\hat\Theta-1}$ \cite{1999ApJ...510..551O}.

As a final note, the expectation value in equation (\ref{eq:c10}), which
derives ultimately from the determinant
in the Gaussian probability density, is small when the noise is small
($C^{\epsilon\epsilon}\ll C^{\Theta\Theta}$) and we are far
from the boundaries of the region of sky surveyed.  This is because
substituting the zero-noise limit for $\hat\Theta$ and
$C^{\hat\Theta\hat\Theta}$ into the expectation value converts it into

\begin{equation}
\Lambda[\hat\Phi] \left< \left[ ((C^{\Theta\Theta})^{-1}\Theta) \nabla
\Theta \right] \right>.
\label{eq:c13}
\end{equation}
We next note that for a statistically isotropic unlensed $\Theta$ and an
all-sky survey, the expectation value in equation
(\ref{eq:c13})  
must vanish because it is a 2-vector (i.e. a vector on $S^2$) and hence a
nonzero value would pick out a preferred
direction.  Near a boundary of the surveyed region, this argument fails
because the boundary breaks rotational
symmetry.  The expectation value in (\ref{eq:c10}) thus acquires a
nonzero value only in the presence
of instrument noise and boundary effects.  Conceptually, we understand
this as a property of
equation (\ref{eq:chat}): noise adds
the $C^{\epsilon\epsilon}$ term to $C^{\hat\Theta\hat\Theta}$, while
boundary effects alter the unit determinant of
$\Lambda$.  Without these effects, $\det C^{\hat\Theta\hat\Theta}=\det
C^{\Theta\Theta}={\rm const}$, and the
expectation value in equation (\ref{eq:c10}), which is merely a
derivative of the log-determinant,
vanishes.
We further note that modes with large noise ($C^{\epsilon\epsilon}\gg
C^{\Theta\Theta}$) do not contribute to $G$ because of the
$C^{\hat\Theta\hat\Theta-1}$ which appears twice in equation
(\ref{eq:c10}).  Since most CMB experiments have only a small range of
$l$ for which $C^{\epsilon\epsilon}$ and $C^{\Theta\Theta}$ are of the
same order, and it is only in this regime and near
boundaries
that the expectation value in equation (\ref{eq:c10}) is important, we
will neglect the expectation value in the remainder of this
paper.  That is, we approximate:

\begin{equation}
G[\Phi] \approx  \nabla\cdot \left[ (C^{\hat\Theta \hat\Theta}[\hat \Phi]^{-1}
\hat\Theta) \Lambda[\hat\Phi] \nabla C^{\Theta\Theta}
\Lambda[\hat \Phi]^{-1} (C^{\hat\Theta \hat\Theta}[\hat \Phi])^{-1} \hat\Theta
\right].
\label{eq:c13b}
\end{equation}

\subsection{\label{sec:s3c}Linearized Version of MLE}

In order to connect equation (\ref{eq:c11}) to previous work on quadratic
estimators, we approximate the right hand side of the
equation to linear order in $\Phi$: $G\approx G_0 + F\Phi$, where the
likelihood gradient $G$ is computed from equation
(\ref{eq:c13b}), $F$ is the matrix of second derivatives $\delta^2{\cal
L}/\delta\Phi\delta\Phi$ (independent of $\Phi$ in the Gaussian approximation),
and $G_0$ is equal to $G$ evaluated with no lensing:

\begin{equation}
G_0 \approx \nabla\cdot \left[ \hat\Theta
(C^{\Theta\Theta}+C^{\epsilon\epsilon})^{-1} \nabla C^{\Theta\Theta}
(C^{\Theta\Theta}+C^{\epsilon\epsilon})^{-1} \hat\Theta \right].
\label{eq:c14a}
\end{equation}
In order to obtain a quadratic (rather than merely a rational) estimator,
the approximate curvature matrix $F$ must be taken
independent of $\hat\Theta$.  We will therefore replace it by its
expectation value $\langle F\rangle$ averaged over $\hat\Theta$.  This
expectation value is:

\begin{equation}
F_{mn}\approx \langle F_{mn}\rangle|_{\Phi} = \int {\cal D}\hat\Theta
P({\hat\Theta}|\Phi) F_{mn}
= \int {\cal D}\hat\Theta e^{-{\cal L}(\Phi|\hat\Theta)} {\delta^2{\cal
L}(\Phi|\hat\Theta)\over\delta\Phi_m\delta\Phi_n},
\label{eq:parts1}
\end{equation}
where $m$ and $n$ are the matrix indices of the matrix of second
derivatives.  We may compute the last integral by requiring that the
probability distribution for
$\hat\Theta$ be properly normalized:

\begin{equation}
1 = \int {\cal D}\hat\Theta P({\hat\Theta}|\Phi) = \int {\cal
D}\hat\Theta e^{-{\cal L}(\hat\Theta|\Phi)}.
\label{eq:parts2}
\end{equation}
Taking the second derivative of this equation with respect to $\Phi$ gives:

\begin{equation}
0 = -\int {\cal D}\hat\Theta e^{-{\cal L}(\hat\Theta|\Phi)}
{\delta^2{\cal L}(\Phi|\hat\Theta)\over\delta\Phi_m\delta\Phi_n}
+ \int {\cal D}\hat\Theta e^{-{\cal L}(\hat\Theta|\Phi)} {\delta{\cal
L}(\Phi|\hat\Theta)\over\delta\Phi_m} {\delta{\cal
L}(\Phi|\hat\Theta)\over\delta\Phi_n},
\label{eq:parts3}
\end{equation}
which enables us to rewrite equation (\ref{eq:parts1}) as:

\begin{equation}
F_{mn} \approx \int {\cal D}\hat\Theta e^{-{\cal L}(\hat\Theta|\Phi)}
{\delta{\cal L}(\Phi|\hat\Theta)\over\delta\Phi_m} {\delta{\cal
L}(\Phi|\hat\Theta)\over\delta\Phi_n}
= \langle G_m G_n \rangle.
\label{eq:parts4}
\end{equation}
Thus the matrix of second derivatives is simply the covariance of the
likelihood gradient.  (Note: we will call the matrix of second
derivatives the Fisher matrix even
though the technical definition of the Fisher matrix differs from $F$ for
a non-Gaussian likelihood function.)  We choose to evaluate $F$ at
$\Phi=0$ (no lensing) for
convenience, although within the Gaussian approximation $F$ can be
evaluated anywhere.  At $\Phi=0$, equation (\ref{eq:c13b}) for $G$
simplifies dramatically and we have:

\begin{equation}
F\approx \langle GG^T\rangle|_{\Phi=0} \approx \left<
\nabla\cdot \left[ (C^{\hat\Theta \hat\Theta-1}
\hat\Theta) \nabla C^{\Theta\Theta}
 C^{\hat\Theta \hat\Theta-1} \hat\Theta
\right]
\left\{ \nabla\cdot \left[ (C^{\hat\Theta \hat\Theta-1}
\hat\Theta) \nabla C^{\Theta\Theta}
 C^{\hat\Theta \hat\Theta-1} \hat\Theta
\right]\right\} ^T
\right>,
\label{eq:c14b}
\end{equation}
which is recognizable as a four-point correlation function of
$\hat\Theta$.  If we switch to the flat sky approximation, and assume
the noise is isotropic, $C^{\Theta\Theta}$ and $C^{\epsilon\epsilon}$
become diagonal in Fourier space.  Then we can compute the
four-point correlation function for Gaussian $\hat\Theta$ using Wick's theorem:

\begin{equation}
\langle \hat\Theta_{{\bf l}_1} \hat\Theta_{{\bf l}_2} \hat\Theta_{{\bf
l}_3} \hat\Theta_{{\bf l}_4} \rangle
=
C^{\hat\Theta\hat\Theta}_{{\bf l}_1} C^{\hat\Theta\hat\Theta}_{{\bf l}_2}
\delta_{{\bf l}_1+{\bf l}_3,0} 
\delta_{{\bf l}_2+{\bf l}_4,0}
+
C^{\hat\Theta\hat\Theta}_{{\bf l}_1} C^{\hat\Theta\hat\Theta}_{{\bf l}_2}
\delta_{{\bf l}_1+{\bf l}_4,0} 
\delta_{{\bf l}_2+{\bf l}_3,0}
+
C^{\hat\Theta\hat\Theta}_{{\bf l}_1} C^{\hat\Theta\hat\Theta}_{{\bf l}_3}
\delta_{{\bf l}_2+{\bf l}_3,0} 
\delta_{{\bf l}_1+{\bf l}_4,0}.
\label{eq:c14c}
\end{equation}
This gives the result for $F$:

\begin{equation}
F_{\bf L} = {1\over 8\pi^2} \int_{{{\bf l}_1}+{{\bf l}_2}={\bf L}}
d^2{{\bf l}_1}
{\left[ {\bf L}\cdot ({\bf l}_1C^{\Theta\Theta}_{l_1} + {\bf
l}_2C^{\Theta\Theta}_{l_2})\right] ^2
\over (C^{\Theta\Theta}_{l_1}+C^{\epsilon\epsilon}_{l_1})
(C^{\Theta\Theta}_{l_2}+C^{\epsilon\epsilon}_{l_2}) },
\label{eq:c15}
\end{equation}
Equation (\ref{eq:c15}) is recognizable (apart from a factor of $4\pi/L^2$
due to normalization
convention) as the noise variance and
optimal weighting derived by Hu \cite{2001ApJ...557L..79H}.  We use the
flat-sky approximation only to compute the noise
curves in Figure \ref{fig:fckappa}; in our simulations we
will evaluate $F$ via a Monte Carlo technique (see Section \ref{sec:s5b}).

We can then construct the maximum likelihood estimator for $\Phi$ under
this approximation,

\begin{equation}
\hat\Phi_{\rm MLE} = -F^{-1}G_0 = -F^{-1}
\nabla\cdot \left[ \hat\Theta
(C^{\Theta\Theta}+C^{\epsilon\epsilon})^{-1} \nabla C^{\Theta\Theta}
(C^{\Theta\Theta}+C^{\epsilon\epsilon})^{-1} \hat\Theta \right],
\label{eq:c16}
\end{equation}
and the corresponding approximate mode of the posterior probability density:

\begin{equation}
\hat\Phi = -(( C^{\Phi\Phi})^{-1} + F)^{-1}G_0 = -(( C^{\Phi\Phi})^{-1} +
F)^{-1}
\nabla\cdot \left[ \hat\Theta
(C^{\Theta\Theta}+C^{\epsilon\epsilon})^{-1} \nabla C^{\Theta\Theta}
(C^{\Theta\Theta}+C^{\epsilon\epsilon})^{-1} \hat\Theta \right].
\label{eq:c17}
\end{equation}
Both of these are recognizable as quadratic estimators, i.e. they are
second-order polynomials in $\hat\Theta$.  By
spherical symmetry, if $\hat\Theta$ is statistically isotropic then the
vector quantity in brackets, and hence
$\hat\Phi$, will
have expectation value zero.  Thus
equations (\ref{eq:c16}) and (\ref{eq:c17}) are measuring the deviation
of $\hat\Theta$ from statistical isotropy that
arises from lensing by a potential $\Phi$.  These deviations from
statistical isotropy in position-space appear as
correlations between different spherical harmonic modes in harmonic
space; see \cite{2001ApJ...557L..79H} for the
associated harmonic-space estimator.  It can be shown
\cite{2001ApJ...557L..79H} that within the linear approximation,
equation (\ref{eq:c16}) provides an unbiased estimate for the lensing
potential $\Phi$ when averaged over an ensemble of
primary CMB anisotopies $\Theta$.

Having determined these approximations, we consider the conditions of
their validity.  The linearization of the r.h.s. of equation
(\ref{eq:c7}) clearly corresponds to a
Gaussian approximation to the likelihood function, with the second-order
Taylor expansion of $\cal L$ carried out around $\Phi=0$.  This can be
expected to be valid when
the maximum likelihood point is ``near'' $\Phi =0$ in the sense that
$\Phi\ll {\cal L}''/{\cal L}'''$ (where the $'$ denotes a functional
derivative with respect to
$\Phi$).  Therefore it would be reasonable to expect that the estimators
in equations (\ref{eq:c16}) and (\ref{eq:c17}) break down when the
lensing effects become large,
i.e. when $C^{\Phi\Phi}$ becomes sufficiently large.  We analyze this
possibility analytically in Sections \ref{sec:s3d} and \ref{sec:s3f} and
numerically in Section
\ref{sec:s6}.

\subsection{\label{sec:primaryrec}Reconstructing the Primary CMB}

We next wish to reconstruct the primary (unlensed) CMB $\Theta$ from
observations $\hat\Theta$ of the lensed
temperature field.  Because our instrument gives us one function on the
sky, $\hat\Theta$, it is not in general possible to simultaneously
reconstruct the primary CMB
anisotropy,
$\Theta$, and the lensing potential $\Phi$.  If the CMB and lensing power
spectra are given, however, we can use the power spectra as a prior and
construct a Bayesian
posterior probability distribution for $\Theta$ and $\Phi$ and maximize
it.  While this does not permit determination of the primary anisotropy
to arbitrary accuracy, it
is the best that one can hope for if only the lensed CMB temperature is
available.  The determination of the lensing potential and primary CMB power
spectra is discussed
in Sections \ref{sec:s4b} and \ref{sec:psp}.

To estimate the primary CMB anisotropy $\Theta$, we take a Gaussian prior
for both the primary CMB and the lensing potential.  This gives
us a joint posterior probability distribution for $\Theta$ and $\Phi$ of
$e^{-\cal R}$, where $\cal R$ is given (up to an additive constant) by

\begin{equation}
{\cal R}[\Theta,\Phi] = {1\over 2} \Theta^T (C^{\Theta\Theta})^{-1} \Theta
+ {1\over 2} \Phi^T (C^{\Phi\Phi})^{-1} \Phi
-\ln P(\hat\Theta |\Theta,\Phi),
\label{eq:rec1}
\end{equation}

where $P(\hat\Theta|\Theta,\Phi)$ is the conditional probability of
observing temperature $\hat\Theta$ given a primary CMB temperature
$\Theta$ and lensing potential $\Phi$.  It is readily noted that $P$ is
simply the instrument noise curve, which we take to be Gaussian:

\begin{equation}
-\ln P(\hat\Theta|\Theta,\Phi) = {1\over 2} (\hat\Theta -
\Lambda[\Phi]\Theta)^T (C^{\epsilon\epsilon})^{-1}
(\hat\Theta - \Lambda[\Phi]\Theta).
\label{eq:rec2}
\end{equation}

Equations (\ref{eq:rec1}) and (\ref{eq:rec2}) formally express the joint
posterior probability distribution for $\Theta$ and $\Phi$.  In order to
reconstruct the primary CMB, we integrate out the lensing potential to
find the negative log posterior probability distribution $\bar{\cal R}$
for $\Theta$:

\begin{equation}
e^{-\bar{\cal R}[\Theta]} = \int {\cal D}\Phi e^{-{\cal R}[\Theta,\Phi]}
=\int {\cal D}\Phi \exp \biggl[ -{1\over 2} (\hat\Theta -
\Lambda[\Phi]\Theta)^T C^{\epsilon\epsilon-1}
(\hat\Theta - \Lambda[\Phi]\Theta)-{1\over 2} \Theta^T C^{\Theta\Theta-1}
\Theta
- {1\over 2} \Phi^T C^{\Phi\Phi-1} \Phi \biggr]
\label{eq:rec3}.
\end{equation}
This equation is difficult to evaluate.  
In the linear approximation, however, we may replace
$\Lambda[\Phi]\Theta$ with $\Theta + \nabla\Phi\cdot\nabla\Theta$; this
makes the integral Gaussian, so it can be evaluated analytically to give

\begin{eqnarray}
\bar{\cal R}[\Theta] = && -{1\over 2}{\cal
G}[\Theta]^T(C^{\Phi\Phi-1}+{\cal F}[\Theta])^{-1}{\cal G}[\Theta]
+{1\over 2}\Theta^TC^{\Theta\Theta-1}\Theta
\nonumber\\*
&& + {1\over 2}(\hat\Theta-\Theta)^T C^{\epsilon\epsilon-1} (\hat\Theta-\Theta)
+ {1\over 2}\ln\det (C^{\Phi\Phi-1} + {\cal F}[\Theta]),
\label{eq:rec4}
\end{eqnarray}
where

\begin{equation}
{\cal G}[\Theta]({\bf x}) = \nabla\cdot \left[
(C^{\epsilon\epsilon-1}(\hat\Theta-\Theta)) \nabla\Theta\right]({\bf x})
\label{eq:rec4a}
\end{equation}
and
\begin{equation}
{\cal F}[\Theta]({\bf x},{\bf y}) = (\nabla_{\bf x}\cdot)(\nabla_{\bf
y}\cdot) (\nabla\Theta({\bf x}) C^{\epsilon\epsilon-1}({\bf x},{\bf y})
\nabla\Theta({\bf y})).
\label{eq:rec4b}
\end{equation}
Note that we have used integration by parts to write ${\cal F}[\Theta]$
and ${\cal G}[\Theta]$.
The matrix ${\cal F}[\Theta]$ is manifestly symmetric and can be seen to
have all nonnegative eigenvalues as follows: if we take any real map
$X({\bf x})$, then

\begin{equation}
X^T{\cal F}[\Theta]X =(\nabla\Theta\cdot\nabla X)^T C^{\epsilon\epsilon-1}
(\nabla\Theta\cdot\nabla X).
\label{eq:rec5}
\end{equation}
Since the inverse noise matrix $C^{\epsilon\epsilon-1}$ is symmetric and
positive-definite, this quantity must be
nonnegative.  Indeed, this can only be zero if $\nabla\Theta\cdot\nabla
X=0$ everywhere, that is, if $X$ is constant on
flows of $\nabla\Theta$.  If we have all-sky coverage and $\Theta$ is
well-behaved, then all of the flow curves of
$\nabla\Theta$ connect at the maxima, minima, and saddle points of
$\Theta$, consequently in this case ${\cal F}$ is
positive-definite except for the constant $l=0$ mode.  Consequently, the
matrix $C^{\Phi\Phi-1}+{\cal F}[\Theta]$ must
be positive-definite, which is required for (\ref{eq:rec4}) to make sense
as a probability distribution (this was also
implicitly assumed in doing the Gaussian integral).  We next define the
(not symmetric!) matrix $H[\Theta]$ by

\begin{equation}
H[\Theta]X = \nabla\Theta\cdot\nabla X
\label{eq:rec8}
\end{equation}
so that ${\cal F}[\Theta] = H[\Theta]^T C^{\epsilon\epsilon-1}
H[\Theta]$.  We can see, using integration by parts, that ${\cal
G}[\Theta] = H[\Theta]^T C^{\epsilon\epsilon-1} (\hat\Theta-\Theta)$.

To make further progress, we use equation (\ref{eq:rec8}) to rewrite the
first term on the right-hand side of equation (\ref{eq:rec4}):

\begin{eqnarray}
\bar{\cal R}[\Theta] = &&
-{1\over 2}(\hat\Theta-\Theta)^T  \left(
C^{\epsilon\epsilon}H[\Theta]^{T-1} C^{\Phi\Phi-1}
H[\Theta]^{-1}C^{\epsilon\epsilon} + C^{\epsilon\epsilon} \right)^{-1}
(\hat\Theta-\Theta)\nonumber\\* &&
+{1\over 2}\Theta^T(C^{\Theta\Theta})^{-1}\Theta
+ {1\over 2}(\hat\Theta-\Theta)^T C^{\epsilon\epsilon-1} (\hat\Theta-\Theta)
+ {1\over 2}\ln\det (C^{\Phi\Phi-1} + {\cal F}[\Theta]).
\label{eq:rec4r}
\end{eqnarray}
Even this equation is too complicated to be useful in this form, so we
will make the replacements $H[\Theta]\rightarrow H[\hat\Theta]$ and
$F[\Theta]\rightarrow F[\hat\Theta]$.  This converts equation
(\ref{eq:rec4}) into a Gaussian posterior probability distribution.  The
peak of the posterior probability distribution is

\begin{eqnarray}
\Theta_{\rm PEAK} = && \left[ C^{\epsilon\epsilon-1} - \left(
C^{\epsilon\epsilon}H[\hat\Theta]^{T-1} C^{\Phi\Phi-1}
H[\hat\Theta]^{-1}C^{\epsilon\epsilon} +
C^{\epsilon\epsilon} \right)^{-1} + C^{\Theta\Theta-1}
\right]^{-1}\nonumber\\* &&
\left[ C^{\epsilon\epsilon-1} - \left(
C^{\epsilon\epsilon}H[\hat\Theta]^{T-1} C^{\Phi\Phi-1}
H[\hat\Theta]^{-1}C^{\epsilon\epsilon} + C^{\epsilon\epsilon}
\right)^{-1} \right] \hat\Theta,
\label{eq:rb}
\end{eqnarray}
and its covariance (i.e. inverse-curvature) is

\begin{equation}
{\rm Cov}[\Theta] = \left[ C^{\epsilon\epsilon-1} - \left(
C^{\epsilon\epsilon}H[\hat\Theta]^{T-1} C^{\Phi\Phi-1}
H[\hat\Theta]^{-1}C^{\epsilon\epsilon} + C^{\epsilon\epsilon}
\right)^{-1} + C^{\Theta\Theta-1} \right]^{-1}.
\label{eq:rc}
\end{equation}

It is instructive to compare equation (\ref{eq:rb}) to other means of
estimating $\Theta$.  Note that determination of $\Theta$ is a nontrivial
task since both lensing and instrument noise must be taken into
account.  In the limit that lensing is negligible
($C^{\Phi\Phi}\rightarrow 0$), we derive $\Theta_{\rm PEAK}\approx
(C^{\epsilon\epsilon-1}+C^{\Theta\Theta-1})^{-1}C^{\epsilon\epsilon-1}
\hat\Theta$, which is recognizable as a simple Wiener filter of
$\hat\Theta$.  In the opposite limit, where instrument noise is
negligible compared to the effects of lensing
(i.e. $C^{\epsilon\epsilon}\rightarrow0$), we derive

\begin{equation}
C^{\epsilon\epsilon-1} - \left( C^{\epsilon\epsilon}H[\hat\Theta]^{T-1}
C^{\Phi\Phi-1} H[\hat\Theta]^{-1}C^{\epsilon\epsilon} +
C^{\epsilon\epsilon} \right)^{-1} \approx H[\hat\Theta]^{T-1}
C^{\Phi\Phi-1} H[\hat\Theta]^{-1}
\label{eq:rd}
\end{equation}
via a first-order Taylor expansion in $C^{\Phi\Phi-1}$.  Substituting
this into equation (\ref{eq:rb}) yields

\begin{equation}
\Theta_{\rm PEAK} = \left( H[\hat\Theta]^{T-1} C^{\Phi\Phi-1}
H[\hat\Theta]^{-1} + C^{\Theta\Theta-1} \right)^{-1}
 H[\hat\Theta]^{T-1} C^{\Phi\Phi-1} H[\hat\Theta]^{-1}  \hat\Theta,
\label{eq:re}
\end{equation}
which is recognizable as a Wiener-filtered temperature map with
$H[\hat\Theta] C^{\Phi\Phi} H[\hat\Theta]^T$ playing the
role of the noise covariance.  This is not surprising since
$H[\hat\Theta] C^{\Phi\Phi} H[\hat\Theta]^T$ is the
covariance of the temperature change due to lensing,
$\tilde\Theta-\Theta$, and under our assumptions the correlation
between $\tilde\Theta-\Theta$ and $\Theta$ vanishes.
Further simplification is possible by noting that, for zero noise, the
likelihood
gradient $G_0$ of Section \ref{sec:s3c} may be written as $G_0=
-H[\hat\Theta]^TC^{\Theta\Theta-1}\hat\Theta $.  Then
the Fisher matrix of Section \ref{sec:s3c} can be approximated as

\begin{eqnarray}
F = \langle G_0G_0^T\rangle_{\Phi=0} = \langle H[\Theta]^T
C^{\Theta\Theta-1} \Theta\Theta^T C^{\Theta\Theta-1}
H[\Theta]\rangle
&& = \langle H[\Theta]^T C^{\Theta\Theta-1} \langle \Theta\Theta^T\rangle
C^{\Theta\Theta-1} H[\Theta]\rangle
\nonumber\\* &&
= \langle H[\Theta]^T C^{\Theta\Theta-1} H[\Theta]\rangle
\approx H[\Theta]^T C^{\Theta\Theta-1} H[\Theta].
\label{eq:rf}
\end{eqnarray}
(The last equality on the first line is justified as follows: since
$H[\Theta]$ is a linear function of $\Theta$, and $\Theta$ is a Gaussian
random field, $\Theta$ and $H[\Theta]$ are jointly Gaussian.  Thus the
expectation value of the four-point function $H[\Theta]^T
C^{\Theta\Theta-1} \Theta\Theta^T C^{\Theta\Theta-1}
H[\Theta]$ can be expanded using Wick's theorem as a sum of three terms,
each of which is a product of two-point functions.  Since $\langle
G_0\rangle=0$, the final expression on the first line of equation
(\ref{eq:rf}) is the only nonvanishing term.)
If we further assume that the lensing effect can be treated as a
perturbation on the background CMB -- an assumption that we have made
already through the linear approximation -- we can approximate
$H[\hat\Theta]\approx H[\Theta]$, which allows equation (\ref{eq:re}) to
be rewritten as

\begin{eqnarray}
\Theta_{\rm PEAK} = H[\hat\Theta] \left( C^{\Phi\Phi-1} +F \right)^{-1}
 C^{\Phi\Phi-1} H[\hat\Theta]^{-1}  \hat\Theta
&& = \hat\Theta - H[\hat\Theta] \left( C^{\Phi\Phi-1} +F \right)^{-1}
 F H[\hat\Theta]^{-1}  \hat\Theta\nonumber\\* &&
= \hat\Theta - H[\hat\Theta] \left( C^{\Phi\Phi-1} +F \right)^{-1}
 H[\hat\Theta]^T C^{\Theta\Theta-1}  \hat\Theta.
\label{eq:rg2}
\end{eqnarray}
Using equations (\ref{eq:c14a}) and (\ref{eq:rec8}) and integration by
parts we see that
$H[\hat\Theta]C^{\Theta\Theta-1}\hat\Theta=-G_0$.
Also, comparison of equation (\ref{eq:rec8}) to the lensing operator
definition, equation (\ref{eq:lambda}), indicates
that in the linear approximation, $\Lambda[\Phi_1]\Theta_1 = \Theta_1 +
H[\Theta_1]\Phi_1$.  This allows us to
simplify equation (\ref{eq:rg2}) to

\begin{equation}
\Theta_{\rm PEAK}
=\hat\Theta + H[\hat\Theta] \left( C^{\Phi\Phi-1} + F\right)^{-1} G_0
= \Lambda[\left( C^{\Phi\Phi-1} +F \right)^{-1} G_0 ]\hat\Theta.
\label{eq:rg}
\end{equation}
This is the observed temperature map ``corrected'' for lensing using the
Wiener-filtered potential map, equation (\ref{eq:c17}).  It is thus the
temperature analogue of the approach used in \cite{2002PhRvL..89a1303K}
for reconstructing primary polarization.

\subsection{\label{sec:s3d}Onset of Nonlinearity}

We examine the validity of the linear approximation leading to equations
(\ref{eq:c16}) and (\ref{eq:c17}) using the
real-space Taylor expansion of the lensing formula, equation
(\ref{eq:lambda}):

\begin{equation}
\tilde\Theta = \Lambda\Theta = \Theta + \nabla\Phi\cdot\nabla\Theta
+ {1\over 2}\nabla\Phi\nabla\Phi :\nabla\nabla\Theta + O(\Phi^3).
\label{eq:lin}
\end{equation}
Quadratic estimator was constructed based on the first-order (i.e. order
$\Phi^1$) effect of lensing on
$C^{\hat\Theta\hat\Theta}$, which neglects the second-order and higher
terms in equation (\ref{eq:lin}), as well as the
covariance of the first-order term.  Thus we expect that this
approximation will be good if the ratio of successive
terms $R$ in equation (\ref{eq:lin}) is small.  As a simple (and
naive!) first approach to determining when the linear
approximation is valid, we note that if $L$ denotes the typical multipole
of $\Phi$, and $l$ denotes the typical
multipole of $\Theta$, then $R=Ll\Phi$.  Since the mean square value of
$\Phi$ is roughly $L^2C^{\Phi\Phi}_L$, we find
that $R^2\approx L^4C^{\Phi\Phi}_Ll^2$, so the linear-order approximation
breaks down at
$C^{\Phi\Phi}>L^{-4}l^{-2}$.  Given that $L^4C^{\Phi\Phi}_L$ has a
maximum of approximately $10^{-6}$, we would then
conclude that nonlinear effects could become important at $l>1000$,
i.e. Planck ($l_{\rm max}\approx 1600$) and
higher-resolution experiments might be susceptible to these effects.

A more refined version of this analysis would examine the covariance
$C^{\hat\Theta\hat\Theta}({\bf x},{\bf y})$ of the observed temperature
instead of simply the
temperature
fluctuation.  This is because for the long-wavelength lensing
($\Phi$) modes, the second-order ($\Phi^2$) corrections to the covariance
are significantly less than
calculated by the naive method above.  Conceptually, one can understand
this by noting that, because the primary CMB is statistically isotropic,
$C^{\hat\Theta\hat\Theta}$ is sensitive to the relative, not absolute,
deflection of photon trajectories.  In the flat-sky approximation we have:

\begin{eqnarray}
C^{\hat\Theta\hat\Theta}({\bf x},{\bf y})
= &&
C^{\epsilon\epsilon}({\bf x},{\bf y}) +
 C^{\Theta\Theta}({\bf x}-{\bf y})
+ (\nabla\Phi({\bf x}) - \nabla\Phi({\bf y}) ) \cdot
\nabla C^{\Theta\Theta}({\bf x}-{\bf y})\nonumber\\* &&
+ {1\over 2} (\nabla\Phi({\bf x}) - \nabla\Phi({\bf y}) )(\nabla\Phi({\bf
x}) - \nabla\Phi({\bf y}) ) :
\nabla\nabla C^{\Theta\Theta}({\bf x}-{\bf y})
+O(\Phi^3).
\label{eq:correxp}
\end{eqnarray}
We note that if $\nabla\Phi$ is slowly varying compared to the separation
of ${\bf x}$ and ${\bf y}$ (that is,
$L\gamma\ll 1$ where $\gamma=|{\bf x}-{\bf y}|$), a near-cancellation
occurs between the linear terms in equation
(\ref{eq:correxp}).  This cancellation reduces the squared expansion
parameter $R^2$ from $L^4C^{\Phi\Phi}_Ll^2$ to
$\gamma^2L^6C^{\Phi\Phi}_Ll^2$.  Since $\gamma$ can take on a wide range
of values from the scale of the lensing mode, $L^{-1}$, down to the limit
of the instrument's
resolution $l^{-1}_{\max}$,
 it is not at all clear how to proceed analytically with this approach.

\subsection{\label{sec:s3f}Bias of quadratic estimator}

Another way to measure the importance of nonlinear terms is to compute
the bias in the quadratic estimator [equation
(\ref{eq:c16})] for $\Phi$, over an ensemble of primary CMB anisotropies
$\Theta$ and instrument noises $\epsilon$ with
the same
lensing potential $\Phi$.  This
bias vanishes in the linear approximation \cite{2001ApJ...557L..79H}.  
It can be computed by noting that the expectation value of $\Phi_{\rm
eq(\ref{eq:c16})}$ is a linear combination of covariance matrix elements of
$C^{\hat\Theta\hat\Theta}$.  We first switch to working in Fourier modes
on a flat sky; in Fourier
space, the two-mode correlation function of the observed temperature is
given by the Fourier transform of
equation (\ref{eq:correxp}):

\begin{eqnarray}
\langle\hat\Theta_{{\bf l}_1}\hat\Theta_{{\bf l}_2}\rangle = &&
(C^{\epsilon\epsilon}_{{\bf l}_1}+C^{\Theta\Theta}_{{\bf
l}_1}) \delta_{{\bf l}_1+{\bf l}_2,0}
+ {1\over\sqrt{4\pi}} ({\bf l}_1+{\bf l}_2)\cdot({\bf l}_1C_{l_1}+{\bf
l}_2C_{l_2}) \Phi_{{\bf l}_1+{\bf l}_2}
\nonumber\\*
&& + {1\over 8\pi} \sum_{{\bf k}_1} \Phi_{{\bf k}_1} \Phi_{{\bf k}_2} \left[
({\bf k}_1\cdot{\bf l}_1)({\bf k}_2\cdot{\bf l}_1) C^{\Theta\Theta}_{l_1} +
({\bf k}_1\cdot{\bf l}_2)({\bf k}_2\cdot{\bf l}_2) C^{\Theta\Theta}_{l_2} -
2 ({\bf k}_1\cdot{\bf J})({\bf k}_2\cdot{\bf J}) C^{\Theta\Theta}_J
\right]
\label{eq:ce2}
\end{eqnarray}
where we have set ${\bf k}_2 = {\bf l}_1+{\bf l}_2-{\bf k}_1$.
Then we may use this two-mode correlation function to evaluate the
expectation values of equations (\ref{eq:c14a}) and hence the
quadratic estimator (\ref{eq:c16}):

\begin{eqnarray}
\left( \langle\Phi_{\rm eq(\ref{eq:c16})}\rangle_{\Theta,\epsilon}
\right) _{\bf L}
= &&\Phi_{\bf L} + {1\over 32\pi^{3/2}F_{\bf L}} \sum_{{\bf k}_1}
\Phi_{{\bf k}_1} \Phi_{{\bf k}_2} 
\nonumber\\
&& \cdot
\sum_{{\bf l}_1}
\Upsilon_{{\bf l}_1,{\bf l}_2}  \left[
({\bf k}_1\cdot{\bf l}_1)({\bf k}_2\cdot{\bf l}_1) C^{\Theta\Theta}_{l_1} +
({\bf k}_1\cdot{\bf l}_2)({\bf k}_2\cdot{\bf l}_2) C^{\Theta\Theta}_{l_2} -
2 ({\bf k}_1\cdot{\bf J})({\bf k}_2\cdot{\bf J}) C^{\Theta\Theta}_J
\right],
\label{eq:bias1}
\end{eqnarray}
where we have set ${\bf l}_2={\bf L}-{\bf l}_1$, ${\bf k}_2={\bf L}-{\bf
k}_1$, and ${\bf J}={\bf k}_1-{\bf l}_1$, and:

\begin{figure}
\includegraphics[angle=-90, width=6in]{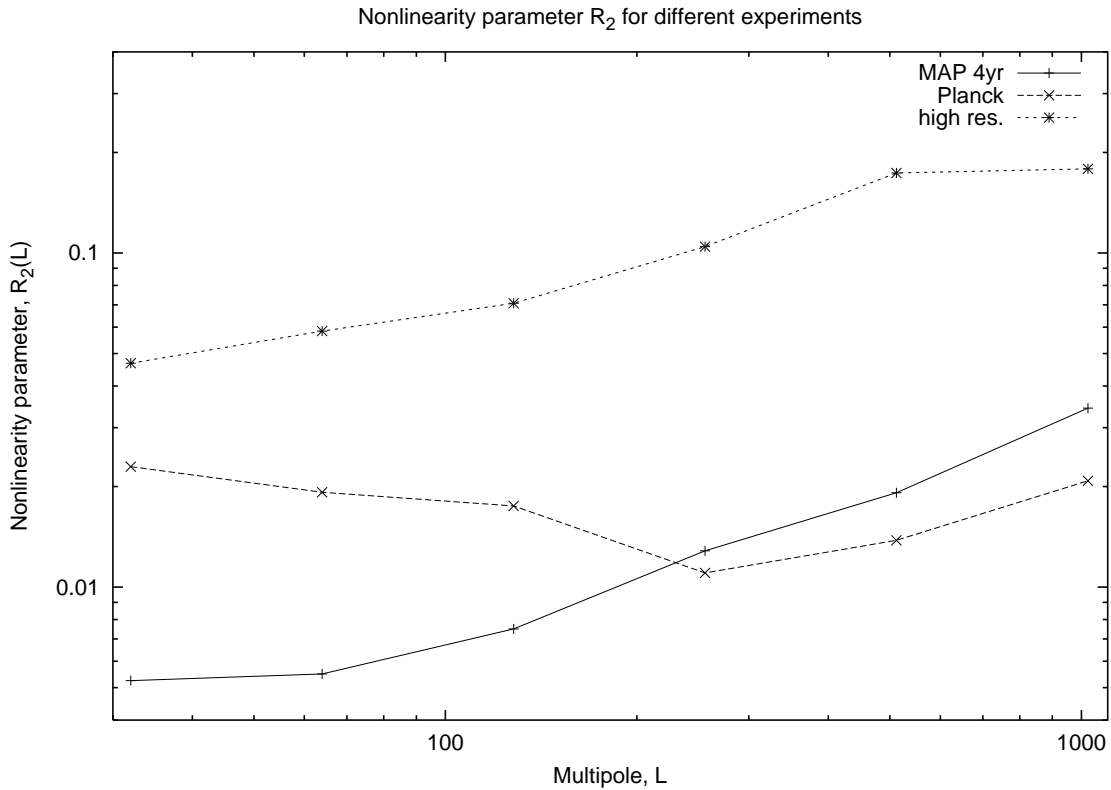}
\caption{\label{fig:fr2}The dimensionless nonlinearity parameter $R_2$,
equal to the ratio of RMS bias in the quadratic lensing potential
estimator [equation
(\ref{eq:c16})] to the RMS
value of the potential,
is plotted here for several experiments as a function of multipole
(wavenumber).  Note that this quantity is less than unity for all of the
experiments.  See Table
\ref{tab:t6} for experiment parameters.}
\end{figure}

\begin{equation}
\Upsilon_{{\bf l}_1,{\bf l}_2} = { {\bf L}\cdot ({\bf
l}_1C^{\Theta\Theta}_{l_1} + {\bf l}_2C^{\Theta\Theta}_{l_2})\over 
(C^{\Theta\Theta}_{l_1}+C^{\epsilon\epsilon}_{l_1})
(C^{\Theta\Theta}_{l_2}+C^{\epsilon\epsilon}_{l_2}) }.
\label{eq:bias2b}
\end{equation}
We can then compute a mean-squared bias by
squaring the bias and ensemble-averaging over $\Phi$.  (Note that since
different Fourier modes of $\Phi$ are uncorrelated, the terms in the sum
over ${\bf k}_1$ will
usually add
incoherently.  The exception to this rule is that terms related by
switching ${\bf k}_1$ and ${\bf k}_2$ are equal, so the mean squared
value of the sum is double the
value obtained by summing the mean square of every term.)  This
mean-squared bias is then given 
by a quadrilateral integral:

\begin{eqnarray}
\left< |\delta\Phi_{\bf L}|^2 \right>_\Phi = &&
\left< \left| \left( \langle\Phi_{\rm
eq(\ref{eq:c16})}\rangle_{\Theta,\epsilon} -\Phi\right)\right| _{\bf L}^2
\right>_\Phi\nonumber\\*
= && {1\over 512\pi^6 F_L^2} \int d^2{\bf k}_1 C^{\Phi\Phi}_{k_1}
C^{\Phi\Phi}_{k_2}
\nonumber\\* &&
 \left\{
\int d^2{\bf l}_1 \Upsilon_{{\bf l}_1,{\bf l}_2} \left[
({\bf k}_1\cdot{\bf l}_1)({\bf k}_2\cdot{\bf l}_1) C^{\Theta\Theta}_{l_1} +
({\bf k}_1\cdot{\bf l}_2)({\bf k}_2\cdot{\bf l}_2) C^{\Theta\Theta}_{l_2} -
2 ({\bf k}_1\cdot{\bf J})({\bf k}_2\cdot{\bf J}) C^{\Theta\Theta}_J
\right] \right\}^2,
\label{eq:bias2}
\end{eqnarray}
We can thus construct a nonlinearity parameter $R_2$ that is the ratio of
the RMS bias to the RMS value of the lensing potential: $R_2^2 = \left<
|\delta\Phi_{\bf
l}|^2 \right>_\Phi/C^{\Phi\Phi}_l$.  This nonlinearity parameter is
plotted as a function of $L$ for three experiments in Figure
\ref{fig:fr2}; the parameters for the
three
experiments -- MAP 4-year data, Planck, and a future high-resolution
experiment -- are shown in Table \ref{tab:t6}.  The $R_2$ nonlinearity
parameter is small
for all but the high-resolution experiment, indicating that the bias in
the quadratic estimator [equation (\ref{eq:c16})] is small.

As a final means of testing the importance of the higher-order terms in
the expansion of $C^{\hat\Theta\hat\Theta}$,
 we conduct a numerical ``experiment'' in Section
\ref{sec:s5} that compares nonlinear [equation (\ref{eq:c12}), without
the small and computationally difficult expectation value] and linear
[equation (\ref{eq:c17})]
lensing potential
estimators.  There we find only modest (10--20\%) improvement in the RMS
error of the lensing potential
reconstruction even for a high-resolution (1 arcminute beam) experiment.

\section{\label{sec:s4}Power Spectrum Estimation}

Having constructed an estimator for the lensing potential $\Phi$, we next
consider its power spectrum
$C^{\Phi\Phi}$.  Conceptually,
the situation here is more complicated because once we average over an
ensemble of lensing potentials derived from the
same power spectrum, the lensed temperature field $\tilde\Theta$ is once
again statistically isotropic with
$\langle\tilde\Theta\tilde\Theta^T\rangle$ diagonal in harmonic
space.  (That is, the off-diagonal elements average to
zero since $\langle\Phi\rangle=0$.)  But we can still construct an
estimator for $C^{\Phi\Phi}=\langle\Phi\Phi^T\rangle$ 
by taking the quadratic estimator for $\Phi$ and computing its
``square''.  The resulting power spectrum estimator is
thus constructed from the four-point correlation function of
$\tilde\Theta$ or (in the presence of
noise) $\hat\Theta$.  It is thus measuring deviations of $\hat\Theta$
from Gaussianity.  We will
show that in the linear approximation, the maximum likelihood estimator
reduces to the quadratic estimator.

We begin this section by formally writing out the likelihood function for
the lensing power spectrum $C^{\Phi\Phi}$ as an integral, and then
approximating this integral as
Gaussian 
 in Section \ref{sec:s4a}.  In Section \ref{sec:s4b}, we approximate the
curvature (inverse-covariance) matrix of this Gaussian in order to obtain
a maximum likelihood 
 estimator that is computationally tractable.  We show in Section
 \ref{sec:s4c} that within the linear approximation, the MLE and quadratic
estimator are
 equivalent.  Computation of the primary CMB power spectrum
 $C^{\Theta\Theta}$ is considered in Section \ref{sec:psp}, and
cross-correlations, e.g. $C^{\Theta\Phi}$, are
considered in Section \ref{sec:s4d}.

\subsection{\label{sec:s4a}Likelihood Function and Gaussian Approximation}

 In principle, we could estimate the power spectrum $C^{\Phi\Phi}_l$ by
 constructing a grand likelihood function $\bar{\cal L}$ given (up to an
additive constant to
 $\bar{\cal L}$, or equivalently a multiplicative constant to
$e^{-\bar{\cal L}}$) by

\begin{equation}
\bar{\cal L} = -\ln\int {\cal D}\Phi P(\Phi|C^{\Phi\Phi}) e^{-\cal L}
 = -\ln\int {\cal D}\Phi \exp \left( -{\cal L}[\Phi]-{1\over 2}
\Phi^TC^{\Phi\Phi-1}\Phi \right)
= -\ln\int {\cal D}\Phi e^{-\cal P},
\label{eq:d1}
\end{equation}
 where the integral over ${\cal D}\Phi$ is a functional integral with the
usual measure on ${\bf R}^N$ where $N$ is the
 number of pixels in the map.  Note that $\bar{\cal L}$ is a function of
the covariance $C^{\Phi\Phi}$; however, we 
 cannot simply maximize the grand likelihood function, equation
(\ref{eq:d1}), and thus obtain an estimate of
 $C^{\Phi\Phi}$, because our map provides us with $N$ real observations
whereas $C^{\Phi\Phi}$ has $N(N+1)/2$ independent
parameters.  In order to
 obtain a meaningful result for the power spectrum, we must restrict the
form of $C^{\Phi\Phi}$.  Fortunately, the
 spherical $SO(3)$ symmetry of the sky provides us with just such a
restriction -- it forces $C^{\Phi\Phi}$ to be
 diagonal in $l$-space.  We will thus assume in this section that
$C^{\Phi\Phi}$ can be written as a linear combination,

\begin{equation}
 C^{\Phi\Phi}({\bf x},{\bf y}) = \sum_\alpha C^{\Phi\Phi}_\alpha {\cal
C}_\alpha ({\bf x},{\bf y}),
\label{eq:d2}
\end{equation}
where the $\alpha$'s are indices labeling the basis covariance functions
and we wish to evaluate the coefficients
 $C^{\Phi\Phi}_\alpha$.  There are two interesting choices of basis
function ${\cal C}_\alpha$.  The first is the 
Legendre polynomials, which span the space of $C^{\Phi\Phi}$ that are
consistent with symmetry requirements.  These
basis functions are given by

\begin{equation}
{\cal C}_l({\bf x},{\bf y}) = {2l+1\over 4\pi} P_l({\bf x}\cdot {\bf y}).
\label{eq:d3}
\end{equation}
This results in coefficients $C^{\Phi\Phi}_l$ that are the power spectrum
of $\Phi$.  The other choice, useful in
the low-SNR case, is to add several functions of the type equation
(\ref{eq:d3}) together to boost the overall SNR, 
i.e. to estimate the lensing power spectrum in a band rather than for
each individual $l$.  In
this case the coefficient $C^{\Phi\Phi}_\alpha$ is a weighted average of
the power spectrum
over the range of $l$ values covered by the basis function ${\cal
C}_\alpha$.

We have now set up the maximum likelihood estimation problem for
$C^{\Phi\Phi}$.  Before proceeding to compute the maximum-likelihood
point, we warn the reader that there is no guarantee that the likelihood
function is devoid of local maxima.  Most of the methods described here
cannot avoid local maxima, nor can they be readily adapted to detect local
maxima.  The exception is the Markov chain method, although the number of
iterations required to escape from a local maximum may be prohibitively large.

Since $N$ is a large number (typically $10^6$--$10^7$), brute-force
integration of equation (\ref{eq:d1}), does not
appear feasible.  There are at least two conceivable approaches to this
problem: a Markov chain (MC) integration, or a
Taylor expansion of the integrand.
While the MC approach is dramatically faster than a brute-force
integration, it is apparent from the high dimensionality
(one dimension for each map pixel) of the problem that many iterations in
the sequence will be necessary for
convergence.  We have not found a computationally feasible implementation
of MC for this problem. The alternate approach
is to Taylor expand $\cal P$ to quadratic order in $\Phi$ around its
minimum $\Phi_{\min}$, i.e. to approximate the
posterior
probability distribution for $\Phi$ as a Gaussian.  This gives

\begin{eqnarray}
{\bar{\cal L}} [C^{\Phi\Phi}_\alpha] =
-\ln\int {\cal D}\Phi e^{-{\cal P}[\Phi, C^{\Phi\Phi}_\alpha ]} && 
= -\ln\int {\cal D}\Phi \exp \left( -{\cal L}[\Phi] - {1\over 2}
\Phi^TC^{\Phi\Phi-1}\Phi \right)
\nonumber\\* &&
\approx
-\ln\int {\cal D}\Phi \exp\left( -{\cal P}[\Phi_{\min}, C^{\Phi\Phi}_\alpha ]
- {1\over2} (\Phi-\Phi_{\min})^T{\delta^2 {\cal P}[\Phi_{\min},
C^{\Phi\Phi}_\alpha ]\over\delta\Phi\delta\Phi}
(\Phi-\Phi_{\min})
\right)
\nonumber\\* &&
\approx {\cal P}[\Phi_{\min}, C^{\Phi\Phi}_\alpha]
+ {1\over 2} \ln\det {\delta^2 {\cal P}[\Phi_{\min}, C^{\Phi\Phi}_\alpha
]\over\delta\Phi\delta\Phi}
\equiv {\cal P}[\Phi_{\min}, C^{\Phi\Phi}_\alpha ] + {1\over 2} \ln\det
K,{\rm ~~~}
\label{eq:d4}
\end{eqnarray}
where the curvature matrix $K$ (the matrix of second derivatives of $\cal
P$ with respect to $\Phi$, evaluated at
$\Phi_{\min}$) has been
introduced and an irrelevant additive constant has been dropped.  Using
equation (\ref{eq:d4}), we seek to minimize the
grand likelihood function $\bar{\cal L}$.  To do this, we differentiate
the final result of equation (\ref{eq:d4}),
yielding

\begin{eqnarray}
0={\partial {\bar{\cal L}}\over\partial C^{\Phi\Phi}_\alpha }
&& = {\partial \over\partial C^{\Phi\Phi}_\alpha }\left(
{\cal P}[\Phi_{\min}, C^{\Phi\Phi}_\alpha ] + {1\over 2} \ln\det
K\right) \nonumber\\
&& = {\partial {\cal P}\over\partial C^{\Phi\Phi}_\alpha }
+ {1\over 2}{\partial\over\partial C^{\Phi\Phi}_\alpha }\ln\det
K|_{\Phi_{\min}}
+ {1\over 2}{\partial \Phi_{\min}^T \over\partial C^{\Phi\Phi}_\alpha }
{\delta\ln\det K\over\delta\Phi}|_{\Phi_{\min}},
\label{eq:d5r}
\end{eqnarray}
where the final (chain-rule) term reflects the shifting position
$\Phi_{\min}$ of the minimum as we change
$C^{\Phi\Phi}$.  There is
no corresponding chain-rule term for ${\cal P}$ because at the minimum,
$\delta{\cal P}/\delta\Phi$ vanishes.  We may
now
evaluate the derivative of ${\cal P}$, noting that only the prior term in
equation (\ref{eq:posterior}) has a dependence
on $C^{\Phi\Phi}$.  Combining the log-determinant of $C^{\Phi\Phi}$ from
the prior with the log-determinant of $K$ in
equation (\ref{eq:d5r}) transforms (\ref{eq:d5r}) into

\begin{equation}
0
= -{1\over 2}\Phi_{\min}^T (C^{\Phi\Phi})^{-1} {\cal C}_\alpha
(C^{\Phi\Phi})^{-1}\Phi_{\min}
+ {1\over 2}{\partial\over\partial C^{\Phi\Phi}_\alpha }\ln\det
(C^{\Phi\Phi}K)|_{\Phi_{\min}}
+ {1\over 2}{\partial \Phi_{\min}^T \over\partial C^{\Phi\Phi}_\alpha }
{\delta\ln\det K\over\delta\Phi}|_{\Phi_{\min}}.
\label{eq:d6}
\end{equation}
At this point
we are confronted with the difficulty of computing the curvature matrix
$K$.  Unfortunately, brute force
computation of $K$ requires $O(N^2)$ computations of $\cal P$, each of
which must require at least $O(N)$ elementary 
operations since it accesses $N$ data points; in practice, a computation
of $\cal P$ involves spherical harmonic transforms consisting of
$O(N^{3/2})$ operations.

\subsection{\label{sec:s4b}Approximating the Curvature Matrix}

In order to compute the curvature matrix in equation (\ref{eq:d6}), we
split it into two parts: the curvature of the likelihood function ($F$,
which is the Fisher matrix in the Gaussian approximation) and the
curvature of the prior, which is always
$(C^{\Phi\Phi})^{-1}$: $K=F+(C^{\Phi\Phi})^{-1}$.  This provides us with
the identity

\begin{equation}
C^{\Phi\Phi}K = C^{\Phi\Phi}F + 1_{N\times N},
\label{eq:d7}
\end{equation}
where $1_{N\times N}$ denotes the $N\times N$ identity matrix.
We use the value of $F$ from the Gaussian approximation: $F=\langle
GG^T\rangle$.  If we are far from boundaries or regions of
nonuniform noise, $F$ is diagonal in harmonic space and we may approximate
it in bins accordingly: 

\begin{equation}
F^{-1}({\bf x},{\bf y}) \approx \sum_\alpha [F^{-1}]_\alpha {\cal
C}_\alpha({\bf x},{\bf y})
\equiv \sum_\alpha {1\over F_\alpha} {\cal C}_\alpha({\bf x},{\bf y}),
\label{eq:d7b}
\end{equation}
where the last equality defines the binned Fisher matrix $F_\alpha$.  In
this approximation, $F=\langle GG^T\rangle$ reduces to

\begin{equation}
F_\alpha = {1\over d_\alpha} \left< G^T{\cal C}_\alpha G\right>,
\label{eq:d8a}
\end{equation}
where the expectation value can be computed by a Monte Carlo analysis, and
$d_\alpha$ is the number of lensing modes in
the band covered by ${\cal C}_\alpha$.  Technically it is best to compute
the Fisher matrix at the value of
$\Phi_{\min}$, however for purposes of computational tractability we only
compute it once at $C^{\Phi\Phi}=0$,
$\Phi_{\min}=0$ [see equation (\ref{eq:e5x})].  In this approximation
$\delta K/\delta\Phi$ vanishes so we will drop the
final term in equation (\ref{eq:d6}).
We can then differentiate the log-determinant of $C^{\Phi\Phi}K$ with
respect to a power spectrum coefficient:

\begin{equation}
{\partial\over\partial C^{\Phi\Phi}_\alpha }\ln\det (C^{\Phi\Phi}K)
= {\rm Tr~} \left[ (C^{\Phi\Phi}F + 1_{N\times N})^{-1} {\cal C}_\alpha F
\right]
= {\rm Tr~} \left[ (C^{\Phi\Phi} + F^{-1})^{-1} {\cal C}_\alpha \right].
\label{eq:d8}
\end{equation}
With this approximation,
(\ref{eq:d6}) and (\ref{eq:d8}) give

\begin{equation}
\Phi_{\min}^T (C^{\Phi\Phi})^{-1} {\cal C}_\alpha
(C^{\Phi\Phi})^{-1}\Phi_{\min}
= {d_\alpha \over
C^{\Phi\Phi}_\alpha + F^{-1}_\alpha},
\label{eq:d9}
\end{equation}
where the denominator in the second term uses the $C^{\Phi\Phi}_l$ value
appropriate for the range of multipoles covered by the $\alpha$ basis function.

\subsection{\label{sec:s4c}Linearization}

If the lensing is sufficiently weak, i.e. if we are in the linear regime
(see Section
\ref{sec:s3d}), and we are only using the $\Phi$'s far from our boundary,
we can solve equation
(\ref{eq:d9}) directly.  To do this, begin by examining equation
(\ref{eq:d9}) in $l$-space (assuming diagonality):

\begin{equation}
\sum_{lm} {|\Phi_{lm}|^2 \over (C^{\Phi\Phi}_\alpha)^2}
= {1 \over
C^{\Phi\Phi}_\alpha + F^{-1}_\alpha},
\label{eq:d11}
\end{equation}
where the sum is over the $d_\alpha$ lensing modes grouped into the band
$\alpha$.  If we now take the multipole moments $\Phi_{lm}=
(C^{\Phi\Phi-1}_\alpha +F_\alpha )^{-1} G_{0lm}$ given by (\ref{eq:c17}),
we derive

\begin{equation}
C^{\Phi\Phi}_\alpha = {1\over d_\alpha F_\alpha^2}
\sum_{lm} |G_{0lm}|^2 - F_\alpha^{-1},
\label{eq:d12}
\end{equation}
which is the same result derived by Hu \cite{2001ApJ...557L..79H} in the
flat-sky approximation.  (It is valid in the
all-sky approximation if we re-interpret $F$ and $G_0$ as all-sky variables.)

\subsection{\label{sec:psp}Primary CMB Power Spectrum}

The grand likelihood function $\bar{\cal L}$ defined in equation
(\ref{eq:d1}) contains the complete dependence of the
probability density of $\hat\Theta$ on the primary CMB power spectrum,
$C^{\Theta\Theta}$.  Thus it can be
simultaneously maximized over $C^{\Phi\Phi}$ and $C^{\Theta\Theta}$.  We
first parametrize the
primary temperature power spectrum $C^{\Theta\Theta}$
in analogy to equation (\ref{eq:d2}):

\begin{equation}
C^{\Theta\Theta}({\bf x},{\bf y}) = \sum_\alpha C^{\Theta\Theta}_\alpha
{\cal W}_\alpha({\bf x},{\bf y}),
\label{eq:p1}
\end{equation}
where the ${\cal W}_\alpha$'s are the basis functions.  Then we
differentiate the Gaussian approximation, equation
(\ref{eq:d5r}), with respect to the coefficients
$C^{\Theta\Theta}_\alpha$ to determine the condition for maximization
of the likelihood $e^{-\bar{\cal L}}$:

\begin{eqnarray}
0={\partial\bar{\cal L}\over\partial C^{\Theta\Theta}_\alpha} && \approx
{\partial\over\partial C^{\Theta\Theta}_\alpha}\left(
 {\cal P}[\Phi_{\min}, C^{\Phi\Phi}_\alpha ] + {1\over 2} \ln\det
K\right) \nonumber\\* &&
= {\partial{\cal P}\over\partial C^{\Theta\Theta}_\alpha}
 + {1\over 2}{\partial\over\partial C^{\Theta\Theta}_\alpha }\ln\det
K|_{\Phi_{\min}}
 + {1\over 2}{\partial \Phi_{\min}^T \over\partial C^{\Theta\Theta}_\alpha
} {\delta\ln\det
K\over\delta\Phi}|_{\Phi_{\min}}.
\label{eq:p2}
\end{eqnarray}
We proceed in analogy to our analysis of the lensing potential power
spectrum in Section \ref{sec:s4b}.
 We neglect the change of $\det K$ with $\Phi$, thus eliminating the last
term in equation (\ref{eq:p2}).  We can
simplify the first term by noting that ${\cal P}$ consists of a prior and
the unmarginalized likelihood ${\cal L}$; the
prior has no dependence on $C^{\Theta\Theta}_\alpha$, while the
unmarginalized likelihood [given by equation
(\ref{eq:lform})] has derivative:

\begin{equation}
 {\partial{\cal L}\over\partial C^{\Theta\Theta}_\alpha} = -{1\over
2}\hat\Theta^T C^{\hat\Theta\hat\Theta-1}
 \Lambda[\Phi] {\cal W}_\alpha \Lambda[\Phi]^T C^{\hat\Theta\hat\Theta-1}
\hat\Theta
 + {1\over 2} {\rm Tr~}\left( C^{\hat\Theta\hat\Theta-1} \Lambda[\Phi]
{\cal W}_\alpha \Lambda[\Phi]^T \right).
\label{eq:p3}
\end{equation}
Combining this with equation (\ref{eq:p2}),
and using equation (\ref{eq:d7}) to eliminate $K$ in favor of $F$ in the
last term, gives us:

\begin{equation}
0= -{1\over 2}\hat\Theta^T C^{\hat\Theta\hat\Theta-1}
 \Lambda[\Phi] {\cal W}_\alpha \Lambda[\Phi]^T C^{\hat\Theta\hat\Theta-1}
\hat\Theta
 + {1\over 2} {\rm Tr}\left( C^{\hat\Theta\hat\Theta-1} \Lambda[\Phi]
{\cal W}_\alpha \Lambda[\Phi]^T \right)
 + {1\over 2} {\rm Tr}\left[ (F+C^{\Phi\Phi-1})^{-1} {\partial
F\over\partial C^{\Theta\Theta}_\alpha} \right].
\label{eq:p4}
\end{equation}
One can readily see that in the absence of lensing, the final term in
this equation vanishes, the $\Lambda[\Phi]$
matrices become the identity, and this equation reduces to the standard
maximum likelihood result for CMB power spectrum
estimation:

\begin{equation}
0= -{1\over 2}\hat\Theta^T C^{\hat\Theta\hat\Theta-1}
{\cal W}_\alpha C^{\hat\Theta\hat\Theta-1} \hat\Theta
+ {1\over 2} {\rm Tr}\left( C^{\hat\Theta\hat\Theta-1} {\cal W}_\alpha \right).
\label{eq:p5}
\end{equation}

\subsection{\label{sec:s4d}Correlation of Lensing With Other Observables}

We may want to compute the correlation of the lensing potential $\Phi$
with some other quantity.  Examples could include
the CMB temperature $\tilde\Theta$, Sunyaev-Zel'dovich or X-ray
observations of hot gases, or galaxy maps.  Since the
focus of this paper is on likelihood methods, and approximations to them,
we will restrict our attention here to the
case of determining $C^{Z\Phi}_l$ where $Z$ is an observable which has a
jointly Gaussian distribution with
$\Phi$.  This situation is expected to be a very good approximation for
the CMB-lensing correlation
$C^{\tilde\Theta\Phi}_l$ introduced by the ISW effect, since ISW is
expected to be apparent primarily on large scales
which are still in the linear regime; some non-Gaussianity in the
potential-induced $\tilde\Theta$ fluctuations may be
expected from nonlinear growth at $l>100$ \cite{2002PhRvD..65d3007V}, but 
this should have negligible effect on the expected signal to noise.  
For the other observables, the situation is complicated by nonlinear
evolution and the method described here should be used with caution.

We will neglect any error in the determination of $Z$.  This is not as
restrictive an assumption as it might seem; if we wish to cross-correlate
$\Phi$ with an observable that has Gaussian error bars, we may write

\begin{equation}
Z = \check Z + \zeta,
\label{eq:x1}
\end{equation}
where $Z$ is the measured value of the observable, $\check Z$ is the
actual value, and $\zeta$ is the error.  If $\zeta$ is Gaussian and
independent of $\check Z$ or $\Phi$, and $\check Z$ is jointly
Gaussian-distributed with the lensing potential $\Phi$, we infer the relations

\begin{equation}
C^{ZZ}_l = C^{\check Z\check Z}_l+C^{\zeta\zeta}_l {\rm\hskip 0.5in
and\hskip 0.5in}
C^{Z\Phi}_l = C^{\check Z\Phi}_l,
\label{eq:x2}
\end{equation}
so that estimating the cross-correlation of $Z$ and $\Phi$ becomes
equivalent to measuring the desired correlation $C^{\check
Z\Phi}_l$.  We can then construct the likelihood function:

\begin{equation}
\bar{\cal L}[C^{\Phi\Phi}_\alpha,C^{Z\Phi}_\alpha] = -\ln \int {\cal
D}\Phi \exp\left[ 
-{\cal L}[\Phi] -{1\over 2}
\left( \begin{array}{c} Z \\ \Phi \end{array} \right)^T
\left( \begin{array}{cc} C^{ZZ} & C^{Z\Phi} \\ C^{\Phi Z} & C^{\Phi\Phi}
\end{array} \right) ^{-1}
\left( \begin{array}{c} Z \\ \Phi \end{array} \right)
\right].
\label{eq:x2a}
\end{equation}

The estimators of Sections \ref{sec:s3b} and \ref{sec:s4b} need only minor
modification in order to do a joint maximum-likelihood analysis of
$C^{\Phi\Phi}$ and $C^{Z\Phi}$.  To see this, note that for a joint
distribution with specified covariance, the expected value of $\Phi$ given
$Z$ is

\begin{equation}
E[\Phi|Z] \equiv \langle\Phi\rangle |_Z = C^{\Phi Z} C^{ZZ-1} Z \equiv AZ,
\label{eq:x3}
\end{equation}
where we have defined the slope matrix $A = C^{\Phi Z} C^{ZZ-1}$.
The variance given $Z$ is

\begin{equation}
C^{\Phi\Phi}|_Z \equiv \left< (\Phi - \langle\Phi\rangle |_Z )^2 \right> =
C^{\Phi\Phi} - C^{\Phi Z}C^{ZZ-1}C^{Z\Phi}, 
\label{eq:x4}
\end{equation}
where we have used $C^{Z\Phi}({\bf x},{\bf y})\equiv \langle Z({\bf
x})\Phi({\bf y})\rangle$.  (Note that $C^{\Phi Z}$ is the
matrix transpose of $C^{Z\Phi}$.)  Equations (\ref{eq:x3}) and
(\ref{eq:x4}) are general for any joint Gaussian distribution, hence
they are valid here even considering the existence of boundaries.  Using
them, we can re-write the likelihood function [equation (\ref{eq:x2a})] as

\begin{equation}
\bar{\cal L}[C^{\Phi\Phi}_\alpha,C^{Z\Phi}_\alpha] = 
{1\over 2}Z^T C^{ZZ-1} Z
-\ln \int {\cal D}\Phi \exp\left\{
-{\cal L}[\Phi] -{1\over 2}(\Phi-E[\Phi|Z])^T [C^{\Phi\Phi}|_Z]^{-1}
(\Phi-E[\Phi|Z])
\right\} ,
\label{eq:x4a}
\end{equation}
which is of the same form as the first (exact!) line of equation
(\ref{eq:d1}).  The additive constant ${1\over 2}Z^T C^{ZZ-1} Z$ has no
effect since we take $Z$ and
$C^Z$ to be constant, so the estimators developed earlier in this paper
to compute $\Phi$ and $C^{\Phi\Phi}$ can be re-written to compute
$\Phi-E[\Phi|Z]$ and
$C^{\Phi\Phi}|_Z$, respectively.  We next construct these estimators
before turning our attention to the problem of estimating the slope
matrix $A$ that relates $Z$ to
$E[\Phi|Z]$.

If we are sufficiently far from a boundary, we
can diagonalize in harmonic space to yield

\begin{equation}
E[\Phi|Z] = \left[ \sqrt{C^{\Phi\Phi}_l\over C^{ZZ}_l} \rho^{Z\Phi}_l
\right] Z = A_l Z
{\rm\hskip 0.5in and\hskip 0.5in}
C^{\Phi\Phi}|_Z = C^{\Phi\Phi} (1-\rho^{Z\Phi2}_l),
\label{eq:x5}
\end{equation}
where $\rho^{Z\Phi}_l$ is the correlation coefficient of the $l$th-order
multipoles.  We note at this point that (assuming $C^{ZZ}$ is known or
has been separately
measured) that the sets of variables $(C^{\Phi\Phi}_l,C^{Z\Phi}_l)$,
$(C^{\Phi\Phi}_l,\rho^{Z\Phi}_l)$, and $(C^{\Phi\Phi}_l |_Z,A_l)$ are
merely different
parametrizations of the same space of models.  We can estimate any of
these pairs; $(C^{\Phi\Phi}_l |_Z,A_l)$ is introduced here precisely
because it is the easiest to
estimate directly.  We can now immediately convert equation
(\ref{eq:c12}), without the trace, to yield 

\begin{equation}
\Phi = E[\Phi|Z] + C^{\Phi\Phi}|_Z G
\label{eq:x6}
\end{equation}
as the mode of the posterior probability distribution [where $G$ is the
likelihood gradient as specified in
(\ref{eq:c14a})].  The power spectrum estimator result, equation
(\ref{eq:d9}), becomes

\begin{equation}
1={C^{\Phi\Phi}_\alpha |_Z + F^{-1}_\alpha \over d_\alpha}
G^T {\cal C}_\alpha G,
\label{eq:x7}
\end{equation}
where $G$ in equation (\ref{eq:x7}) is evaluated at the solution to
equation (\ref{eq:x6}).  These equations specify the
conditions for the likelihood ${\bar{\cal L}}(C^{\Phi\Phi},\rho^{Z\Phi})$
to be stationary with respect to first-order
variations in $C^{\Phi\Phi}|_Z$ with $A_l$ constant.  In order to
complete the analysis, we must also identify the
condition for $\bar{\cal L}$ to be stationary with respect to first-order
variations in the $A_l$ (slope) coefficients
in equation (\ref{eq:x5}) with $C^{\Phi\Phi}|_Z$ constant.  For these
variations, if we again approximate the Fisher matrix as $F=\langle
GG^T\rangle |_{\Phi=0}$, we
derive a constant curvature matrix $K$.  Then, parametrizing the $A_l$ in
bands in analogy to equation (\ref{eq:d2}) gives

\begin{equation}
A({\bf x},{\bf y}) = \sum_\alpha A_\alpha {\cal B}_\alpha({\bf x},{\bf y}),
\label{eq:x8}
\end{equation}
where it is assumed that ${\cal B}_\alpha$ and hence $A$ are
symmetric.  We obtain a maximum likelihood condition on
$A_\alpha$ by differentiating $\bar{\cal L}$:

\begin{equation}
{\partial {\bar{\cal L}}\over\partial A_\alpha} = {\partial {\cal
P}\over\partial A_\alpha}
= {\partial {\cal L}\over\partial A_\alpha}
 = \int_\Omega d^2{\bf x} {\delta {\cal L}\over\delta \Phi({\bf x})}
{\partial \Phi({\bf x})\over\partial A_\alpha}
 = \int_\Omega d^2{\bf x} {\delta {\cal L}\over\delta \Phi({\bf x})}
{\partial E[\Phi |Z]({\bf x})\over\partial A_\alpha}
= Z^T {\cal B}_\alpha G.
\label{eq:x9}
\end{equation}
Note that we have taken $\Phi - E[\Phi |Z]$ to be constant here; this was
merely a convenient choice.  [Since we have
 maximized $\cal P$ with respect to $\Phi$, we can choose any first-order
variation in $\Phi$ without affecting the
 derivative in equation (\ref{eq:x9}).]  It follows that the joint
maximum-likelihood estimator for $(C^{\Phi\Phi},A)$
satisfies

\begin{equation}
Z^T {\cal B}_\alpha G = 0.
\label{eq:x10}
\end{equation}
We can then reconstruct the full lensing power spectrum and
cross-correlation using the relations 

\begin{equation}
C^{\check Z\Phi}_l = C^{ZZ}_lA_l{\rm \hskip 0.5in and\hskip 0.5in}
C^{\Phi\Phi}_l = A_l^2C^{ZZ}_l + C^{\Phi\Phi}_l|_Z. 
\label{eq:x10b}
\end{equation}
If we take the linear approximation of Section \ref{sec:s3d}, that
$G\approx G_0+F\Phi$, and approximate diagonality in harmonic space as in
Section \ref{sec:s4c}, equation (\ref{eq:x10}) becomes:

\begin{equation}
A_\alpha = -{Z^T{\cal B}_\alpha G_0\over Z^T{\cal B}_\alpha F{\cal
B}_\alpha Z}.
\label{eq:x11}
\end{equation}
If we note that $C^{Z\Phi}=C^{ZZ}A_l$, we note that this is the same
result as obtained by correlating the linearized maximum likelihood
estimator, equation (\ref{eq:c16}) with $Z$.  In the case of the
$\tilde\Theta\Phi$ correlation ($Z=\hat\Theta$), which is of interest for
investigating the ISW effect, the numerator is cubic in $\hat\Theta$,
i.e. the maximum likelihood estimator for $A_\alpha$ is the same as that
computed from the bispectrum.

As a final point, we note that for the $\tilde\Theta\Phi$ correlation,
the error $\zeta=\epsilon$ in $Z=\hat\Theta$ is not entirely independent
of the estimation procedure for $\Phi$, since we are after all
determining $\Phi$ from the CMB temperature measurements.  Since we
assumed $\zeta$ to be uncorrelated with $\Phi$ and its determination,
this is a potential flaw in our calculations as applied to the
$\tilde\Theta\Phi$ correlation.  We expect the error induced by this
effect to be small, since the ISW effect is most important on the large
scales where the instrument noise is small: $C^{\epsilon\epsilon}_l\ll
C^{\Theta\Theta}_l$.  We additionally note that the determination of
$\Phi$ primarily uses information from much higher $l$.

\section{\label{sec:s5}Implementation and Results}

\begin{table}
\caption{\label{tab:t6}Reference Parameters for CMB Experiments}
\begin{tabular}{lcccc}
\hline\hline
Experiment & ~~~~~ & $w^{-1/2}$ / 2.725 K radian & ~~~~~ & $\sigma$ /
arcmin \\
\hline
MAP (4 yr) & & $5.6\times 10^{-8}$ & & $13$ \\
Planck & & $2.9\times 10^{-9}$ & & $6$ \\
high-res. & & $5.0\times 10^{-10}$ & & $1$ \\
\hline\hline
\end{tabular}
\end{table}

In order to demonstrate the feasibility of computing the estimators above
in a realistic situation, and to assess their
 performance, we ran several simulations in which a data set was generated
and analyzed.  The data sets are generated on
 a full sphere assuming isotropic Gaussian temperature fluctuations,
lensing potential, and instrument noise.  For Planck
 and the high-resolution reference experiment ($l_{\max}\approx 3500$,
beam FWHM = 1 arcminute), we reconstruct the
lensing
 potential and compare the reconstruction and original map.  The lensing
power spectrum was estimated for the Planck-type
 experiment, but computer time constraints prevented a similar analysis
for the high-resolution experiment.

\begin{figure}
\includegraphics[angle=-90, width=6in]{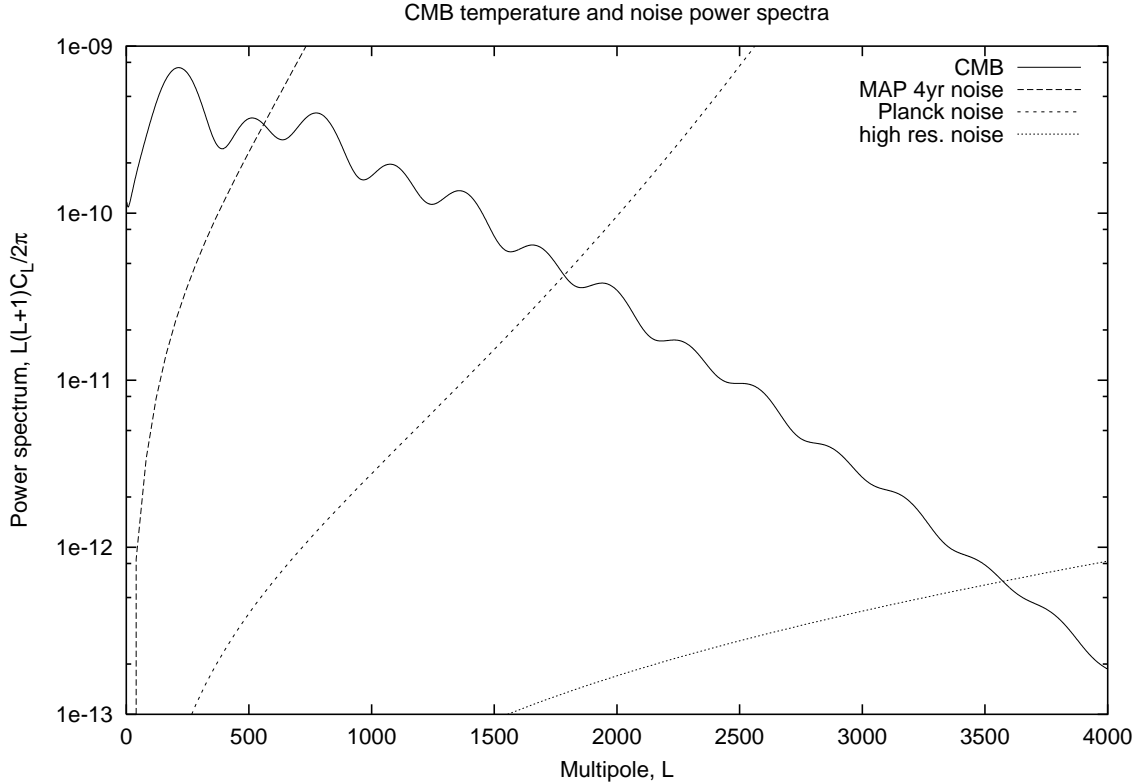}
 \caption{\label{fig:fcmb}The solid line illustrates the model primary CMB
 temperature power spectrum $l(l+1)C_l^{\Theta\Theta}/2\pi$.  The noise
curves
 $l(l+1)C_l^{\epsilon\epsilon}/2\pi$ are shown for MAP 4-year data (top,
long-dashed), Planck (center, short-dashed),
and the high resolution reference experiment (bottom, dotted).}
\end{figure}

\begin{figure}
\includegraphics[angle=-90, width=6in]{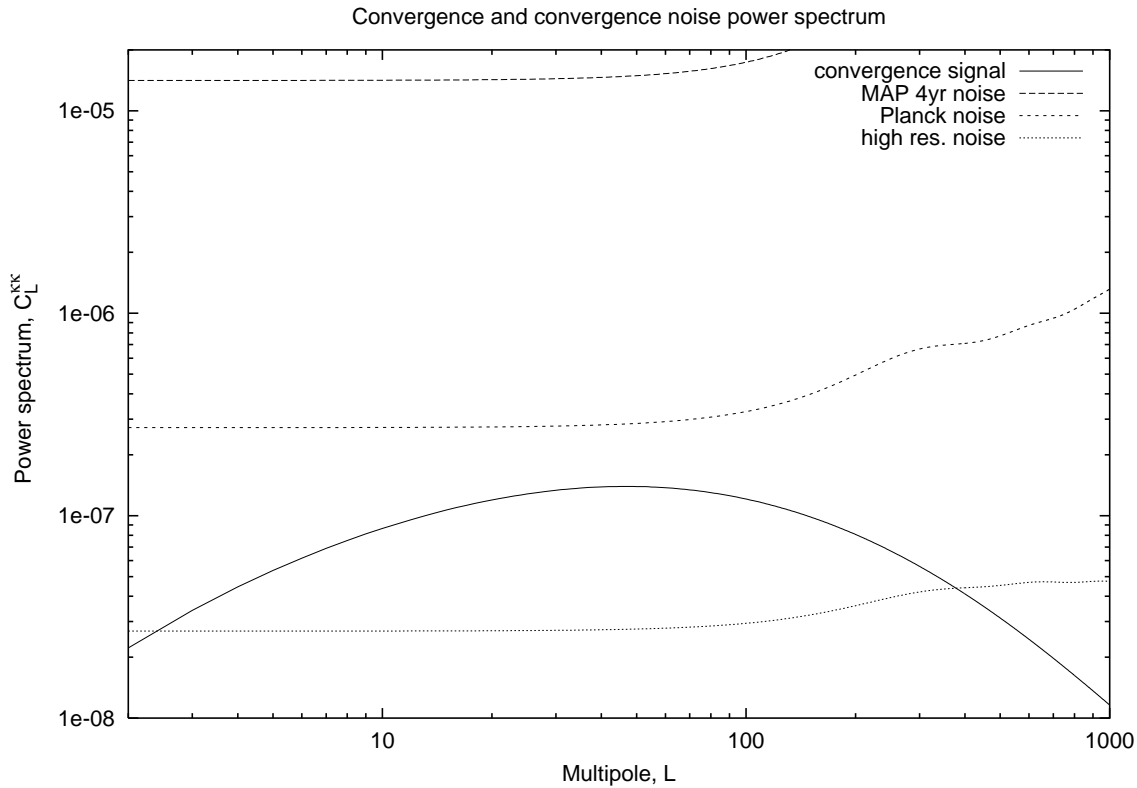}
 \caption{\label{fig:fckappa}The solid line illustrates the model
convergence power spectrum
 $C_l^{\kappa\kappa}=l^2(l+1)^2C^{\Phi\Phi}/4$.  The noise curves
 $l^2(l+1)^2(F_l^{\Phi\Phi})^{-1}/4$ are shown for (top to bottom) MAP
4-year data, Planck, and the
high-resolution
reference experiment, using curvature matrix
elements $F_l$
computed from equation (\ref{eq:c15}).}
\end{figure}

\subsection{\label{sec:s5a}Utilities}

 A lensing simulation requires the capability to work with maps on the
unit sphere, or some subset thereof, particularly
 the capability to perform the elementary algebraic and calculus
operations and to perform convolutions and both forward
 and reverse spherical harmonic transforms (SHT).  We therefore require
the use of a map projection or grid.  In order to
 perform SHT in a reasonable amount of time, we must use an isolatitude
projection, i.e. one in which horizontal lines
 are parallels of the same latitude.  Furthermore, we found conformality
 to be convenient for differentiation and useful for reducing gridding
errors (see
 Section \ref{sec:s5b}).  The only projection with these properties is the
Mercator projection, in which the map
 coordinates $x$ and $y$ are related to the longitude $\phi$ and
colatitude $\theta$ by the formulas $\phi=\tau
(x-x_0)$ and $\cos\theta = \tanh (\tau y)$.

 The conformal magnification $\Gamma$ defined by
 $ds^2=\Gamma^2(dx^2+dy^2)$ satisfies $\Gamma = \tau\sin\theta$.  A map of
some quantity $A$ is stored as a two-dimensonal
 array $A(x,y)$ of values at the points of integer $x$ and $y$.  Spherical
 harmonic transforms are performed by transforming first the longitude
 ($x$ or $\phi$) direction to produce the Fourier coefficients of $A$ at
 constant latitude and then the latitude ($y$ or $\theta$) direction.  On
 a grid with $N$ points, this is an $O(N^{3/2})$ process.  Convolutions
are performed with two successive SHT's.

\subsection{\label{sec:s5b}Estimator for Map of Lensing Potential}

 Our implementation presently approximates the estimator (\ref{eq:c12}) as
 follows.  The expectation value is ignored since it is expected to be
 small; see the discussion following (\ref{eq:c12}).  We must also
approximate the vector

\begin{equation}
 {\bf V} = -\left[ (C^{\hat\Theta\hat\Theta}[\Phi]^{-1} \hat\Theta
) \Lambda[\Phi] \nabla C^{\Theta\Theta}
\Lambda[\Phi]^{-1} (C^{\hat\Theta\hat\Theta}[\Phi])^{-1}\hat\Theta \right],
\label{eq:e2}
\end{equation}
and use it to determine the likelihood gradient $G=\nabla\cdot{\bf
V}$.  (Note that $G={\delta{\cal L}\over\delta\Phi}$ is a scalar function
 on $\Omega$.)  Because of difficulties computing $C^{-1}\hat\Theta$ in a
 reasonable amount of time, we chose to approximate $C^{-1}\hat\Theta$ by
 a sequence of (i) filtering of $\Lambda^{-1}\hat\Theta$ using the
harmonic-space kernel
$C^{\Theta\Theta}/(C^{\Theta\Theta}+C^{\epsilon\epsilon})$ and
(ii) convolution with the $(C_l)^{-1}$ kernel.  Of these steps, both
break down near the boundaries and (ii) breaks down when the lensing is
strong enough so that the noise $C^{\epsilon\epsilon}$ in $\hat\Theta$ is
no longer a good approximation to the noise in
$\Lambda^{-1}\hat\Theta$.  We note that if $C^{\epsilon\epsilon}$ were
flat (i.e. $l^2C^{\epsilon\epsilon}\propto l^2$), this approximation
would become exact far from the boundaries.  (``Real'' instrument errors
show some increase in $C^{\epsilon\epsilon}$ at high $l$ due to finite
beam size \cite{1995PhRvD..52.4307K}.)  In order to reduce gridding
errors, the $(C_l)^{-1}$ operation is performed by convolving with the
kernel $[l(l+1)C_l]^{-1}$ and then taking the Laplacian.
 In order to avoid boundary effects, ${\bf V}$ is multiplied by a function
 $q$ that is equal to one inside $\Omega$ far from the boundary, but falls
off smoothly to zero at the boundary.

 After computing ${\bf V}$, we take its divergence $G=\nabla\cdot{\bf
 V}$; then we must determine $C^{\Phi\Phi}G$.  In order to reduce errors
due to the gridding
 (pixelization), we perform the convolution in two steps.  First, we apply
 an inverse Laplacian operator $\nabla^{-2}$, and then we apply the
 remainder of the convolution, $l(l+1) C^{\Phi\Phi}_l$.  Because we use a
 conformal coordinate system, the inverse Laplacian can be done in the
 plane where gridding errors vanish (the forward and reverse Fourier
 transforms are exact inverses of each other, even on a discrete grid,
 which does not occur for SHT).  This is important since the low-$l$ modes
 of $G$, which correspond to the lensing modes that can be recovered at
 moderate signal-to-noise, are buried in high-$l$ noise due to the power
 spectrum $l^2C^{GG}_l\propto\approx l^5$ in the range of interest $50\le
 l\le 1000$.  The inverse Laplacian operation does not add significant
 time to the computation because it utilizes a fast Fourier transform
 requiring $O(N\log N)$ operations, whereas the computation time is
 dominated by SHT's requiring$O(N^{3/2})$ time.  With some attention paid
to gridding issues, this two-step process may turn out to be unnecessary.

 An iterative procedure is needed for solving equation
(\ref{eq:c12}).  The obvious iterative procedure, $\Phi_{n+1} =
  C^{\Phi\Phi}G[\Phi_n]$, is (in the linear approximation) unstable for
any lensing mode with
SNR$^2=C^{\Phi\Phi}F>2$, and hence is not a good
choice.  We therefore use the underrelaxed version,

\begin{equation}
\Phi_{n+1} = (1-f)\Phi_n + f C^{\Phi\Phi} G[\Phi_n],
\label{eq:e3}
\end{equation}
 where $f$ is a convergence parameter.  In the linear approximation,
convergence would require $0<f<2/(1+{\rm SNR}^2)$,
however, a smaller value
 of $f$ is necessary in practice to avoid instabilities resulting from
boundary effects and nonlinear lensing effects.

\subsection{\label{sec:s5c}Power Spectrum}

 We use equation (\ref{eq:d9}) to estimate the power spectrum
 $C^{\Phi\Phi}$.  The basis functions of choice have constant
$l^2(l+1)^2C_l^{\Phi\Phi}$ within some
 band $l_{\min}\le l\le l_{\max}$.  The number of modes covered by the
basis function ${\cal C}_\alpha$ can be estimated as

\begin{equation}
 d_\alpha =A(\Omega) \sum_{l=l_{\min}}^{l_{\max}} {2l+1\over 4\pi} =
{A(\Omega)\over 4\pi} \left[ (l_{\max }+1)^2-l_{\min
}^2 \right].
\label{eq:e4}
\end{equation}
 The estimator, equation (\ref{eq:d9}), then can be written in the
iterative form:

\begin{equation}
 C^{\Phi\Phi}_\alpha = C^{\Phi\Phi}_\alpha \left[ { C^{\Phi\Phi}_\alpha +
F^{-1}_\alpha \over d_\alpha }
(\nabla^{-2}G)^T P_\alpha \nabla^{-2}G
\right] ^\beta,
\label{eq:e5}
\end{equation}
 where $P_\alpha$ is the projector onto the band $l_{\alpha,\rm min}\le
l\le l_{\alpha,\rm max}$, i.e. the operation that
 filters out all multipoles not included in this band.  We use
 $\nabla^{-2}G$ here because it and its spherical harmonic transform are
 already being computed for the estimation of the map of $\Phi$.  The
 parameter $\beta$ is an adjustable convergence parameter.  The Fisher
matrix $F_\alpha$ is computed by the Monte Carlo procedure:

\begin{equation}
 F_\alpha = {1\over d_\alpha} \left< (\nabla^{-2}G_0)^TP_\alpha
\nabla^{-2}G_0 \right>,
\label{eq:e5x}
\end{equation}
 where the average is taken over an unlensed temperature field (including
noise).

Note that equation (\ref{eq:e5}) exhibits a difference from the quadratic
estimator, equation (\ref{eq:d12}): while 
(\ref{eq:d12}) can, in principle, be negative, $P_\alpha$ is
positive-definite and hence (\ref{eq:e5}) can never yield any result less
than zero.  It is straightforward to show that in this case, assuming the
linearized approximation of Sections \ref{sec:s3c} and \ref{sec:s4c}, and
assuming a positive initial guess is used for the power spectrum to start
the iteration, that (\ref{eq:e5}) tends to zero (estimates no
power).  Because negative results are replaced by zeroes, equation
(\ref{eq:e5}) technically converges to a biased estimator, with
expectation value

\begin{equation}
 {\langle C^{\Phi\Phi}_\alpha({\rm est.})\rangle\over
C^{\Phi\Phi}_\alpha({\rm actual})} = 1 - {1\over 2}{\rm erfc~}{x\over\sqrt 2}
+ {1\over\sqrt{2\pi}}x^{-1}e^{-x^2/2}
= 1+ \pi^{-1/2} e^{-x^2/2} \left( x^{-3} - 3x^{-5} + 5!!x^{-7} - ... \right),
\label{eq:bias}
\end{equation}
where $x=d_\alpha C^{\Phi\Phi}_\alpha F_\alpha /2$ is the squared
signal-to-noise ratio (SNR$^2$) in the power spectrum determination.  If
the signal-to-noise ratio is large ($x\gg 1$) then this bias is
irrelevant.  Note that in the context of maximum likelihood estimation, a
negative power spectrum estimate does not make sense, because the
corresponding probability distribution for $\Phi$ and hence the likelihood
integral, equation (\ref{eq:d1}), are ill-defined.  In particular, the
avoidance of negative power spectra is not an artifact of any
approximation we have made.

Obtaining convergence from the coupled iterative estimators, equations
(\ref{eq:e3}) and (\ref{eq:e5}) requires some care.  Convergence depends
not only on the values of the parameters $f$ and $\beta$, but also on the
pattern of how many times the map is updated using (\ref{eq:e3}) each time
the power spectrum is updated using (\ref{eq:e5}).  As an extreme example,
we note that if $f$ and $\beta$ are taken to be very small, and we
alternate between updating the map and power spectrum, convergence can be
expected only for negative $\beta$; whereas if we iterate the map many
times between iterations of the power spectrum, convergence requires
positive $\beta$.  After some experimentation, we found that iterating the
map many ($M\gg 1/f$) times between iterations of the power spectrum and
taking $\beta=1$ resulted in convergence.

\subsection{\label{sec:s5e}Improvements}

While the implementation described here is sufficient for evaluating the
importance of non-linear effects, much work remains before it could be
used to analyze real  
data.  First, real data have boundaries (if for no other reason than the
presence of a galactic plane cut) and usually have inhomogeneous
noise.  Thus,  
the $C^{-1}$ operation used here will need to be performed by actual
matrix inversion rather than  
by convolution.  The latter also becomes necessary in the event of
non-uniform noise.  Also, the iteration of equation
(\ref{eq:e5}) converges slowly and for  
long-wavelength modes ($1/l$ comparable to the size of the gridded
region) may fail to converge entirely.

Additionally, it would be desirable to use a better approximation to
equation (\ref{eq:d1}) than the Gaussian 
approximation, equation (\ref{eq:d4}), but we were unable to identify a
computationally tractable method of doing this.

\subsection{\label{sec:s6}Results}

\begin{figure}
\includegraphics[angle=-90, width=6in]{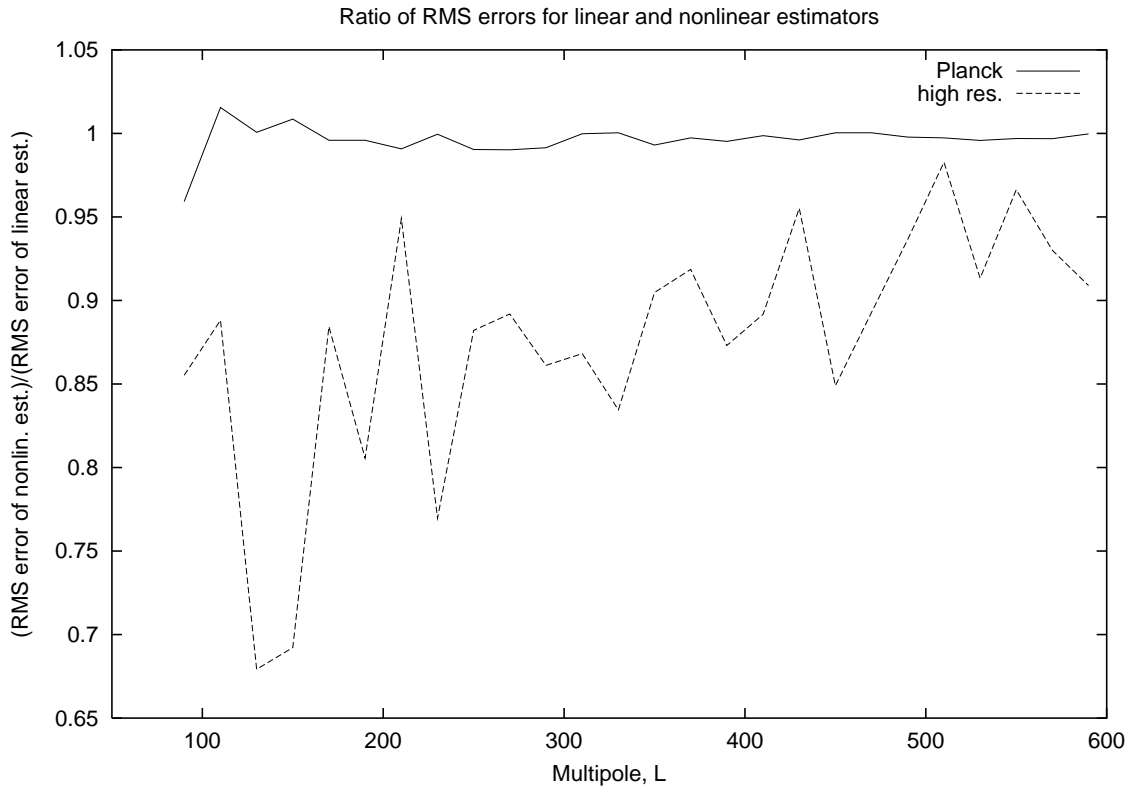}
 \caption{\label{fig:ratios}The ratio of the root mean squared error
($\sqrt{\rm MSE}$) for the nonlinear estimator,
 equation (\ref{eq:e3}), to that of the linear estimator, equation
(\ref{eq:c17}), in bins of $\Delta l$=20.  Results are
 obtained from a Monte Carlo simulation, which is responsible for the
bumpiness of the graph.  The solid line is for
 Planck parameters, and the dotted line is for the high-resolution
experiment (see Table \ref{tab:t6}).}
\end{figure}

Here we investigate the effects of nonlinearity on future CMB
experiments.  We use the form for the instrument noise
\cite{1995PhRvD..52.4307K}:

\begin{equation}
C_l^{\epsilon\epsilon} = w^{-1} e^{l(l+1)\sigma^2/8\ln 2},
\label{eq:six1}
\end{equation}
where the weight $w$ and beam full-width at half maximum $\sigma$ are
parameters, and the beam spot is assumed to be 
 Gaussian.  We use a primary CMB power spectrum $C_l^{\Theta\Theta}$
 generated by {\sc cmbfast}, assuming a flat LCDM universe with parameters
$H_0$=72
km/s/Mpc,
 $T_{\rm CMB}$=2.725 K, $Y_{\rm He}$=0.24, $N_\nu$=3 (massless),
 $\Omega_b$=0.04, $\Omega_{cdm}$=0.30.  The primary CMB model is shown in
 Figure \ref{fig:fcmb}.  The lensing power spectrum, shown in Figure
\ref{fig:fckappa}, is computed normalized to $\sigma_8=1$.

 We compare the linearized estimator to the ``full'' nonlinear estimator
(as implemented here) for two
 experiments: the upcoming Planck satellite mission, and the proposed
Atacama Cosmology Telescope (ACT) as an example  
 of upcoming high resolution, low noise experiments.  The parameters for
the Planck and the
high-resolution reference experiments
 are shown in Table \ref{tab:t6}.  (The MAP 4-year experiment is also
shown for comparison.)  For purposes of
 computational tractability, we have restricted ourselves to a small
portion of the sky: the Planck simulation was run on
 a $751\times 751$ grid with spacing at equator $\tau = 5\times 10^{-4}$
radians (even though Planck is an all-sky
 experiment), and the high-resolution simulation was run on a $1251\times
1251$ grid with $\tau = 1.5\times 10^{-4}$
radians.  (Note the ACT survey region is a long, rectangular
 stripe on the sky as opposed to a more compact patch.  Because of our
implementation's susceptibility to boundary
 effects, we cannot do our simulations on a stripe.)  The solid angles
covered by the simulations are 0.14 sr ($\sim$ 2\%
 of the Planck survey area) for the Planck-type experiment and 0.035 sr
for the high-resolution experiment.

 The results of these simulations are shown in Figure
\ref{fig:ratios}.  The convergence map errors (i.e. $\kappa_{\rm
 est} - \kappa$) for oth the nonlinear estimator, equation
(\ref{eq:e5}) and the linear estimator, equation
 (\ref{eq:c17}) were computed.  The convergence map errors were then
Fourier-transformed (since we are working on a small
 patch of sky), yielding the error amplitude $\kappa_{\bf l}$ for each
Fourier mode.  The modes were then sorted into
 bins of $\Delta l=20$ according to their $l$-value, and an RMS amplitude
$\sqrt{\bar{\kappa_{\bf l}^\ast\kappa_{\bf
 l}}}$ was computed for each bin.  The ratios of these RMS amplitudes are
plotted in Figure \ref{fig:ratios}.  Note that
 for the Planck experiment, there is only a slight advantage in using the
nonlinear estimator, whereas for the
high-resolution
experiment, the accuracy of
 the reconstruction is improved by using the full nonlinear estimator,
equation (\ref{eq:c17}).

 Both the comparison via simulation of the linear and nonlinear estimators
 (Figure \ref{fig:ratios}) and the semi-analytic bias calculation (Figure
\ref{fig:fr2}) are
 methods of assessing the validity of the linear approximation.  Both of
 them suggest that nonlinear effects are more important for the
 higher-resolution experiment than
 for Planck, but (at least for the experiment considered here) are not
 dominant.  Note, however, that for the high-resolution experiment the
semi-analytic calculation found
 nonlinear effects to be more important at higher $l$, whereas the
 simulation found a greater improvement in switching to the nonlinear
estimator at lower $l$.  Note,
 however, that the semi-analytical calculations of Section \ref{sec:s3d}
 and the simulation of this section are not measuring the same
quantity: in Section \ref{sec:s3d} we
 were examining the bias of equation (\ref{eq:c16}), whereas here we are
 considering the mean squared error of the optimally filtered version of
that estimator.

 We also simulated the performance of the linear [equation (\ref{eq:d12})]
and nonlinear [equation (\ref{eq:e5}), 16
 iterations] convergence power spectrum estimators for Planck
parameters.  These were performed on the aforementioned
 0.14 sr patch of sky, for seven $l$ bins: 100--150, 150--200, 200--280,
280--360, 360--440, 440--520, and 520--600.  The
 results are shown in Figure \ref{fig:powerspec}.  The (Monte Carlo) mean
of each estimator, computed from $n=10$ trials
 of area 0.14 sr each, are shown.  Note the similar performance of the
estimators except in the low-$l$ bands.  Note that
 in the full Planck experiment ($\approx$8 sr), the error bars would be
smaller by a factor of $\approx\sqrt{1.4/8}$.

\begin{figure}
\includegraphics[angle=-90, width=6in]{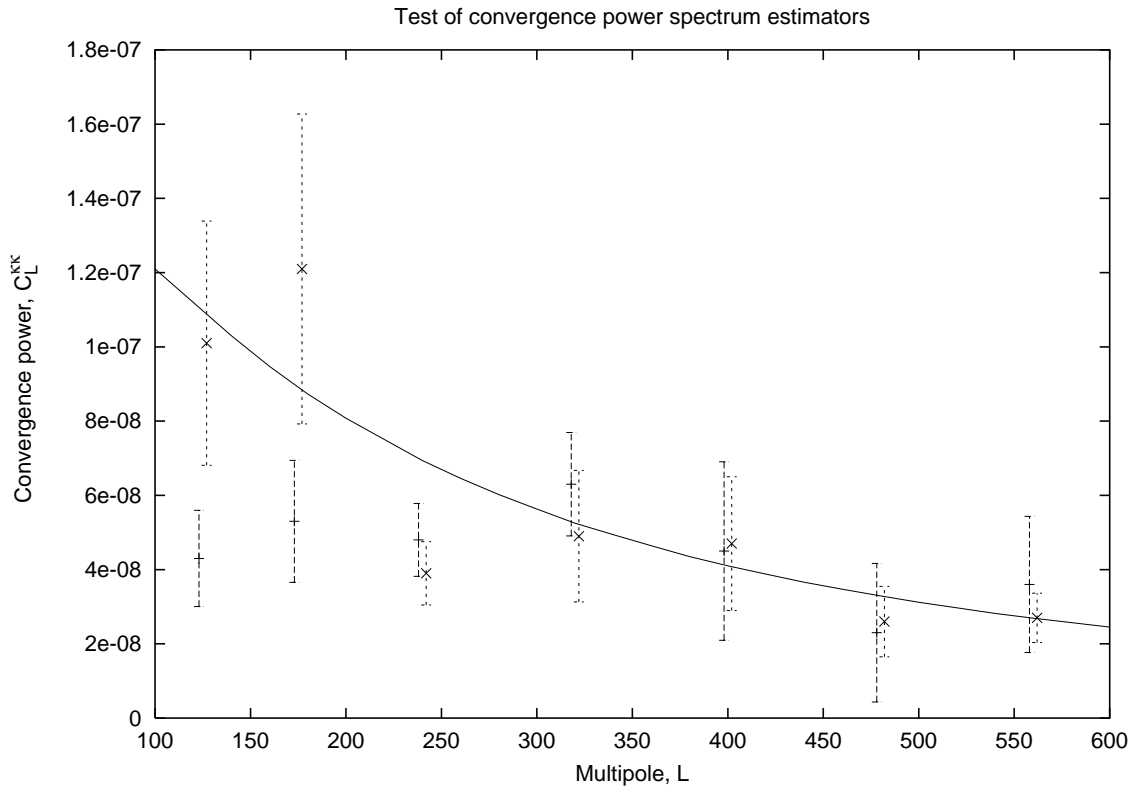}
 \caption{\label{fig:powerspec}The true convergence power spectrum,
 $C^{\kappa\kappa}_l$, is shown by the solid line.  The points (with error
bars) indicate estimated
 convergence power spectra from the linear (+ points) and nonlinear
 ($\times$ points) estimators (equations (\ref{eq:d12}) and
(\ref{eq:e5}) respectively).  To prevent the
error bars from overlapping
 and causing confusion, we have displaced the data point for the linear
 estimator slightly to the left and the data point for the nonlinear
estimator slightly to the
 right.  The error bars are the $1\sigma$ Monte carlo error bars on the
expectation value of the estimator.
 The estimated power spectra plotted are averages over $10$ trials of 0.14
sr solid angle each using Planck parameters,
 and thus shows the error bar on the power spectrum using data from a
region of area $A(\Omega)=1.4$ sr. 
}
\end{figure}

We may test both the linear and nonlinear estimators for bias using the
$t$-test.  The $t$ statistic for band $\alpha$ is given by

\begin{equation}
t[\alpha] = {{\rm SampleMean}(C^{\Phi\Phi}_\alpha) - C^{\Phi\Phi}_\alpha
\over \sqrt {s^2/n}},
\label{eq:ttest}
\end{equation}
where $s^2$ is the sample variance of the $C^{\Phi\Phi}_\alpha$.  The 
$t$-test results are shown in Table \ref{tab:st1}.  A positive $t$
statistic indicates that we are overestimating the power spectrum, a
negative $t$ statistic indicates that we are underestimating it.  The $t$
statistics here are designated $t_9$ in the table because they have 9
degrees of freedom.  Also shown in the table is the two-tailed $p$-value,
i.e. the probability of a perfect $t$ random variable (with 9 degrees of
freedom) having absolute value exceeding $t_9$:

\begin{equation}
 p = 2 \int_{|t_9|}^\infty {\Gamma(n/2+1/2)\over \sqrt{n\pi} \Gamma(n/2)}
\left( 1+{x^2\over n}\right) ^{-(n+1)/2} dx.
\label{eq:studentt}
\end{equation}
If the power spectrum estimator is unbiased and normally distributed, the
 $p$-value for each $l$ bin is uniformly distributed between 0 and
 1.  (Warning: because, for each estimator, we derived the $p$-values for
 all the $l$ bins from the same 10 simulations, there is no reason to
 believe that the $p$-values are independent.)  We note that, for the
 high-$l$ bins ($l>200$), the linear and nonlinear estimators give similar
 results; both are consistent with being unbiased, although the nonlinear
 estimator shows a lower sample variance.  (This is partially the result
 of negative power-spectrum estimates being set to zero by the nonlinear
 estimator.)  In the two lowest-$l$ bins, the sample variance of the
 nonlinear estimator is enormous.  We note that in some of our
 simulations, the nonlinear estimator assigned anomalously large (in one
 case $>4\times 10^{-7}$) values of $C^{\kappa\kappa}$ to these bins; this
 suggests a problem with the estimator.
This may be due to smearing of the bins by the finite width of the
scanned region (width $\approx$0.38 sr) or may
 represent a problem with the iterative procedure (e.g. convergence to a
local maximum of the likelihood).

\begin{table}
\caption{\label{tab:st1}Convergence Power Spectrum Estimators: $t$-test}
\begin{tabular}{ccrcrcrcrcr}
\hline\hline
$l$ range & ~~~~~ & $d_\alpha$ & ~~~~~ & linear $t_9$ & ~~~~~ & linear
$p$ & ~~~~~ & nonlin. $t_9$ & ~~~~~ & nonlin. $p$
\\
\hline
100--150 & & 127 & & -5.08 & & 0.0007 & & -0.23 & & 0.8200 \\
150--200 & & 178 & & -2.23 & & 0.0527 & &  0.79 & & 0.4523 \\
200--280 & & 392 & & -2.16 & & 0.0592 & & -3.66 & & 0.0052 \\
280--360 & & 522 & &  0.73 & & 0.4822 & & -0.19 & & 0.8558 \\
360--440 & & 653 & &  0.17 & & 0.8654 & &  0.31 & & 0.7670 \\
440--520 & & 783 & & -0.51 & & 0.6238 & & -0.71 & & 0.4976 \\
520--600 & & 914 & &  0.49 & & 0.6352 & &  0.03 & & 0.9763 \\
\hline\hline
\end{tabular}
\end{table}

Due to the excessive computation time requirements, we were unable to run
a similar simulation of power spectrum
estimators for the high-resolution experiment.  Such a simulation would
be interesting because the nonlinear lensing
potential estimator showed improvements at the $\ge 10$\% level over the
quadratic estimator for this experiment.

\section{\label{sec:s7}Conclusions}

Weak lensing of CMB temperature maps has been recongized for some time as
a potential probe for mapping the mass
distribution of the universe (in projection), and determining quantities
derivable from such a map: its power spectrum
and cross-correlation with the CMB or other maps.  In the past several
years, methods for carrying out this statistical
analysis have been proposed
\cite{1999PhRvD..60d3504S,1999PhRvL..82.2636S} and dramatically improved
\cite{2001ApJ...557L..79H}.  We have shown, by comparison to
likelihood-based approaches, that quadratic estimator
\cite{2001ApJ...557L..79H} for the lensing potential [equivalent to our
equation (\ref{eq:c16})] is very close to optimal
for the Planck experiment.  That is, for this experiment, there is no
hope of further reduction in the statistical noise
of the lensing potential.  Similarly, simulations of the fully nonlinear
power spectrum estimator 
do not show much improvement over the linear version.

If the lensing potential $\Phi$ can be treated as a Gaussian random
field, and for experiments for which the linearized
approximation suffices, 
then our maximum-likelihood analysis indicates that the CMB temperature
bispectrum and trispectrum are optimal
estimators for the temperature-lensing potential cross-correlation
($C^{\tilde\Theta\Phi}$) and lensing power spectrum,
respectively.  There is, in this case, no additional information in the
fifth-order and higher statistics of the CMB.   
These higher-order statistics may be useful if the lensing potential is
non-Gaussian; for example, the matter density
bispectrum will result in a similar bispectrum in the quadratic
estimator, equation (\ref{eq:c16}), i.e. it its effect
will be seen in the CMB six-point correlation function.  

For the higher-resolution experiments, our results indicate that the
residual error in the lensing
potential maps can be reduced by switching from the quadratic estimator
to a ``full'' likelihood-based estimator.  In
order to make this approach practical, further work will be needed to
develop a version of the algorithm that works near
the survey boundaries and to improve the stability of the
algorithm.  Other approaches to the functional integral in
equation (\ref{eq:d1}) besides the Gaussian approximation used here, such
as Markov chains, could be used.  It is even
possible that, due to use of a better approximation to the integral, the
error can be reduced further.  However, given
that the high-resolution (1 arcminute beam) simulation only showed
improvement in RMS error at the 10--20\% level, it
may be preferable to
simply
use the quadratic estimator for these experiments, accept this minor loss
in signal-to-noise ratio, and avoid the
difficulties associated with the nonlinear estimator.

One problem for both the quadratic estimator approach
\cite{2001ApJ...557L..79H} and our likelihood-based approach are
extremely sensitive to errors in the primary CMB
power spectrum, $C^{\Theta\Theta}_l$.  In the quadratic approach, this
can be seen by noting that the power spectrum $C^{\Phi\Phi}$ is obtained
by differencing two
quantities which may be very close to each other in equation
 (\ref{eq:d12}); the inverse Fisher matrix $F_\alpha^{-1}$ may be tens of
 times greater than $C^{\Phi\Phi}_\alpha$ for Planck (see Figure
 \ref{fig:fckappa}) and consequently the quantities being subtracted in
 (\ref{eq:d12}) may differ by only several percent.  In the likelihood
 approach of Section \ref{sec:s5c} this problem is masked by the formalism
 of the iterative scheme, but it is still there.  Note that this problem
 is more serious for the lower-resolution experiments.  It is apparent
that a lensing power spectrum analysis must be accompanied by extremely
accurate determination of $C^{\Theta\Theta}_l$, or an estimation scheme
must be introduced that is robust against small errors in
$C^{\Theta\Theta}_l$ must be introduced, or both.

We have performed no analysis here of lensing estimators using the CMB
polarization.  In \cite{2002ApJ...574..566H} a quadratic estimator
analysis for the CMB polarization has been performed.  Because the
polarization is a Gaussian random field whose covariance depends on the
lensing potential (i.e. there are observed covariances $C^{\hat\Theta\hat
E}$, $C^{\hat E\hat B}$, etc. analogous to the $C^{\hat\Theta\hat\Theta}$
used here), an analysis analogous to that of Sections \ref{sec:s3} and
\ref{sec:s4} establishes that the quadratic estimator approach is optimal
if the lensing is sufficiently weak.  Because our simulation code cannot
handle polarization, we cannot determine whether realistic lensing is
sufficiently weak or whether a nonlinear maximum-likelihood analysis is
required to make full use of polarization data sets. 
Any high resolution polarization experiment will have 
one of its main goals gravity wave detection in B mode. Since weak 
lensing creates B modes out of E modes \cite{1998PhRvD..58b3003Z} it is
important to 
remove this contamination as well as possible by using the 
weak lensing reconstruction \cite{2002PhRvL..89a1303K,2002PhRvL..89a1304K}. 
Given that polarization and its E and 
B decomposition is sensitive to the direction of polarization in addition 
to its amplitude, it may be more susceptible to the errors induced 
by the linearization procedure. 
In this case the nonlinear analysis 
will be essential to exploit fully the potential of any future high 
resolution CMB polarization experiment. 

\begin{acknowledgments}
The authors wish to thank Matias Zaldarriaga, Nikhil Padmanabhan, Lyman
Page and Vassilios Papathanakos for useful
comments and help.
CH is supported through the NASA Graduate Student Researchers Program,
grant NASA GSRP-02-OSS-079. US is 
supported by NASA ATP and LTSA grants, 
NSF CAREER grant and grants from
David and Lucille Packard Foundation and Alfred P. Sloan Foundation.
\end{acknowledgments}

\appendix*

\section{\label{sec:aphi}Potentials, Convergence, and Projected Density}

Here we sketch a derivation of equation (\ref{eq:potform}), and use this
to relate the lensing potential $\Phi$ to quantities of more direct
physical interest.  We use a Robertson-Walker metric with a Newtonian
perturbation (i.e. a weak perturbation $|\Psi |\ll 1$ induced by
nonrelativistic matter):

\begin{equation}
ds^2 = a^2 \left[ -(1-2\Psi)d\tau^2 + (1+2\Psi)(dr^2 + S(r)^2
d\omega^2) \right] 
\label{eq:ap1}
\end{equation}
where $a$ is a function of the conformal time $\tau$, and $\Psi$ is the
gravitational potential, generally a function of all the coordinates.  The
comoving distance is $r$, and $\omega\in S^2$ is a direction on the unit
sphere with the usual line element $d\omega^2$.  We have used the sinelike
function $S(r) = k^{-1/2}\sin (k^{1/2}r)$, and will use its derivative,
the cosinelike function $C(r) = \cos (k^{1/2}r)$, and their ratio
$T(r)=S(r)/C(r)$, where $k$ is the spatial curvature.  We use as an
initial condition $\tau=0$ at present, and normalize $a(\tau=0)=1$.  The
simplest way to find the photon deflection is to consider the conformal
metric (which must have exactly the same null geodesics):

\begin{equation}
d\tilde s^2 = -(1-4\Psi)d\tau^2 + dr^2 + S(r)^2 d\omega^2
\label{eq:ap2}
\end{equation}
In this metric we compute for the null geodesics (to linear order in
$\Psi$ and assuming that the geodesic is nearly radial):

\begin{equation}
{d^2\omega\over dr^2} = {C(r)\over S(r)} {d\omega\over dr} -
2{\partial\Psi\over\partial\omega}
\label{eq:ap3}
\end{equation}
In the Born approximation, where the gradient over $\omega$ is evaluated
along the unperturbed line-of-sight, this is an inhomogeneous linear
equation in ${\bf n}$ which can be solved by the Green's function method
to find $\omega$ at the last scattering surface.  The result is that the
null geodesic arriving at ``us'' ($r=\tau=0$) from direction ${\bf n}$ is
found to have originated in direction ${\bf n}+\nabla\Phi({\bf n})$,
where:

\begin{equation}
\Phi({\bf n}) = -2 \int_0^{r_{ls}} dr \Psi(r{\bf n},-r) \left( {1\over
T(r)} - {1\over T(r_{ls})} \right) 
\label{eq:ap4}
\end{equation}
Through the use of the trigonometric identities and their hyperbolic
counterparts we can show that: 

\begin{equation}
{1\over T(r)} - {1\over T(r_{ls})} = {S(r_{ls} - r)\over S(r)S(r_{ls})}
\label{eq:ap4b}
\end{equation}
with which equation (\ref{eq:ap4}) can be shown to be equivalent to the
forms provided by, e.g. \cite{2001ApJ...557L..79H,2000PhRvD..62d3007H}.

Since the lensing potential $\Phi$ is represented here as a projected
gravitational potential, it would make sense that its second derivative
$\kappa = -{1\over 2}\nabla^2\Phi$ would represent a projected density
perturbation.  This is indeed the case, although there are other
contributions to $\kappa$.  If we define $\Delta$ to be the comoving
three-dimensional Laplacian (i.e. on $dr^2 + S(r)^2 d\omega^2$), as
distinguished from the two-dimensional Laplacian $\nabla^2$ on the unit
sphere, we have the usual relation for $\Delta$:

\begin{equation}
 \Delta = {1\over S(r)^2}\left[ \nabla^2 + {\partial\over\partial r}
\left( S(r)^2 {\partial\over\partial r} \right) \right]
\label{eq:ap5}
\end{equation}
If we solve this relation for $\nabla^2$, we can split $\kappa=-{1\over
2}\nabla^2\Phi$ into two terms: one involving the $\Delta$ operator and
one involving the radial operator:

\begin{equation}
\kappa = \int_0^{r_{ls}} dr \left( {1\over T(r)} - {1\over T(r_{ls})}
\right) \biggl[ S(r)^2 \Delta\Psi
- {\partial\over\partial r}\left( S(r)^2 {\partial\Psi\over\partial r}
\right) \biggr]
\label{eq:ap6}
\end{equation}
The first term can be replaced with a density using Poisson's equation,
thus generating a projected density.  To study 
the second term, we replace the partial derivative over $r$ (at constant
$\tau$) with a total derivative along the line 
of sight and a time derivative.  The time derivative can be neglected here
if the matter is nonrelativistic.  [Indeed we 
have already made this assumption implicitly when we write equation
(\ref{eq:ap1}).]  Next integrate by parts so that 
the $d/dr$ acts on $1/T(r)-1/T(r_{ls})$.  (Since $S(0)=0$, the surface
terms generated by the integration by parts will 
vanish.)  Then we use the identity ${d\over dr}{1\over T(r)}=-1/S(r)^2$ to
convert equation (\ref{eq:ap6}) into: 

\begin{equation}
\kappa ({\bf n})
=4\pi G_N\int_0^{r_{ls}} dr \left( {1\over T(r)} - {1\over T(r_{ls})} \right)
S(r)^2 a(-r)^2\delta\rho(r{\bf n},-r)
-\Psi(r_{ls}{\bf n},-r_{ls}) + \Psi(0)
\label{eq:ap7}
\end{equation}
where $G_N$ is the universal gravitation constant and $\delta\rho$ is the
density perturbation.  Note that 
convergence can be broken into two components: a component due to the
density fluctuations $\delta\rho$ along the line 
of sight, and a component due to the potential difference between the
source and observer.  The second component is due 
to
tidal forces acting to separate the trajectories of CMB
photons; conceptually, it has the same origin as the 
compression of the sky into a small solid angle near the zenith as seen by
an observer near a black hole 
despite the absence of any mass-energy along the line of sight.

Finally, we attempt to determine the magnitude of the potential-difference
contribution to the convergence.  Here 
$\Psi(0)$ is a
constant (isotropic) and can be removed by a gauge transformation, so we
do not consider it further.  The term 
$\Psi(r_{ls})$ can be estimated based on the CMB temperature fluctuation
using the Sachs-Wolfe relation $\Psi(r_{ls}) = 
3\Theta$ \cite{1967ApJ...147...73S}; thus the contribution of the
$\Psi(r_{ls})$ term to $\kappa$ is $-3\Theta$.  Since 
$C^{-3\Theta,-3\Theta}_l=9C^{\Theta\Theta}_l$ is always at least a factor
of $l^2$ less than the overall power spectrum 
$C^{\kappa\kappa}_l$, its effect on the convergence power spectrum is
subdominant with respect to cosmic variance. 

\bibliography{cosmo,cosmo_preprints}

\end{document}